\documentclass[aip,prl,preprint]{revtex4-1}
\usepackage{graphicx}
\usepackage{amsmath}
\usepackage{amssymb} 
\usepackage[usenames, dvipsnames]{color}

\begin{document}

\title{Simulations of Coulomb systems confined by polarizable surfaces using periodic Green functions}

\author{Alexandre P. dos Santos}
\email{alexandre.pereira@ufrgs.br}
\affiliation{Instituto de F\'isica, Universidade Federal do Rio Grande do Sul, Caixa Postal 15051, CEP 91501-970, Porto Alegre, RS, Brazil.}
\affiliation{Fachbereich Physik, Freie Universit\"at Berlin - 14195 Berlin, Germany.}

\author{Matheus Girotto}
\email{matheus.girotto@ufrgs.br}
\affiliation{Instituto de F\'isica, Universidade Federal do Rio Grande do Sul, Caixa Postal 15051, CEP 91501-970, Porto Alegre, RS, Brazil.}

\author{Yan Levin}
\email{levin@if.ufrgs.br}
\affiliation{Instituto de F\'isica, Universidade Federal do Rio Grande do Sul, Caixa Postal 15051, CEP 91501-970, Porto Alegre, RS, Brazil.}

\begin{abstract}

We present an efficient approach for simulating Coulomb systems confined by planar polarizable surfaces. The method is based on the solution of Poisson equation using periodic Green functions. It is shown that the electrostatic energy arising from surface polarization can be decoupled from the energy of periodic replicas. This 
allows us to combine an efficient Ewald summation method for the replicas with the polarization contribution calculated using Green function techniques.  We apply the method to calculate density profiles of ions confined between charged dielectric and metal interfaces.

\end{abstract}

\maketitle

\newpage
\section{Introduction}

Efficient simulations of charged systems are of fundamental importance for physics, chemistry, and biology.
Because of the long range nature of the Coulomb force one can not use simple periodic boundary conditions
which are sufficient for systems with short range interactions.  Instead one is forced to construct an
infinite set of replicas of the original system, so that a particle in the main simulation cell interacts with all the other particles in the cell, as well as with all the periodic replicas.  To efficiently sum over the replicas of the system Ewald summation methods have been developed~\cite{AlTi87,Fr02,Ewa21,PeLe95,YoPe93,KoPe92,DeCh98,DeCh982}. Originally, Ewald summation was used to calculate the bulk energy of ionic crystals and, in particular, the Madelung constant. 
Ewald summation is based on the separation of the Coulomb potential into long and short range contributions.  The short range part can be treated using the usual periodic boundary conditions, while the long range part can be efficiently summed in the Fourier space.   Unfortunately, the method loses much of its usefulness when the full 3d symmetry is broken, which is the case when interfaces are present. This is due to appearance of special functions in the two dimensional Fourier transform, leading to slow convergence of the lattice sums~\cite{Lek91,WiAd97,Maz05}. This notwithstanding, there are many important systems with a broken symmetry: ionic liquids at electrified interfaces~\cite{FeKo14,LaMa07,JiWu16,KoOs16,FeIv17,DoLe17}, charged nanopores~\cite{WoMu07,HaCo14,FeKo11}, nanoconfined electrolytes~\cite{DoLe15,DoLe16,KlWu15}, just to cite a handful of examples. These systems can present new phenomena, such as like-charged attraction~\cite{LiLo99,HaLu10,SaTr11,NeOr00} and charge reversal~\cite{AlPe11,NgSh02,WaWu17}, which are hard to describe analytically~\cite{Lev02}, hence the importance of fast simulation methods. To overcome the difficulty of using 2d Ewald summation, a number of approaches have attempted to extend the efficient $3$d Ewald summation method to systems with slab geometry~\cite{YeBe99,KaMi01,ArDe02,DoLe16}. These approaches rely on the introduction of a sufficiently large vacuum region between the undesired replicas to diminish their interaction in non-periodic direction.  To account for the conditional convergence of the lattice sums, the Ewald summation must be performed in a ``plane-wise" manner, leading to an additional correction to the usual 3d Ewald energy. The method was shown to be very efficient for simulating systems with reduced symmetry.  The difficulty, however, arises when the simulation cell
is bounded by the polarizable surfaces such as metal electrodes or phospholipid membranes.  If there is only
one polarizable surface present, it is straightforward to extend the techniques described above using the usual image charge construction~\cite{NaHe11,WaMa16,GiDo16,JiCr13,BaDo11,DoBa11,DiDo12}. However, if the simulation cell is bounded by two polarizable surfaces, the situation becomes much more difficult since the image  construction results in an infinite set of image charges.  Therefore, both metallic~\cite{SiSp95,LaMa07} and dielectric confinements~\cite{DoLe16,GiLe16,ZwCr13,SoCr12} make simulations substantially more difficult. A common procedure relies on the calculation of the induced surface charge at the interfaces using minimization of the electrostatic energy or using the discontinuity of displacement field~\cite{RoRo13,ChRo14,ZwLa15,HeEi04,XuLu15,KaHo10}. This makes the simulations very slow, restricting the system size to small number of particles. Recently, we~\cite{DoLe172} introduced an approach that does not rely on energy minimization, but is restricted to metal plates only. If the dielectric contrast is not too large, dos Santos and Levin~\cite{DoLe15} showed that it is possible to sum over the infinite set of image charges. The rate of convergence, however, deteriorates with the dielectric contrast, restricting the range of applicability of this method. There are also other approaches in the literature to deal with polarizable surfaces based on Lekner-like summation~\cite{LeHo13,ArHo08,ArHo07}. Every approach has its own advantages and disadvantages.

In the present paper we will introduce a general method for calculating electrostatic energy of Coulomb systems confined by planar polarizable surfaces, either metallic or dielectric. The method is based on the exact solution of Poisson equation~\cite{Jac99} using periodic Green functions with either Dirichlet or Neumann boundary conditions. The advantage of the new
method is that it is very fast and easy to implement.  A standard 3d Ewald summation code can, therefore, be easily adopted 
to study confined Coulomb systems in slab geometry. As an application, we will calculate the density profiles of ions confined between charged dielectric and metal surfaces.

\section{Green function}

\begin{figure}[h]
\begin{center}
\includegraphics[scale=0.40]{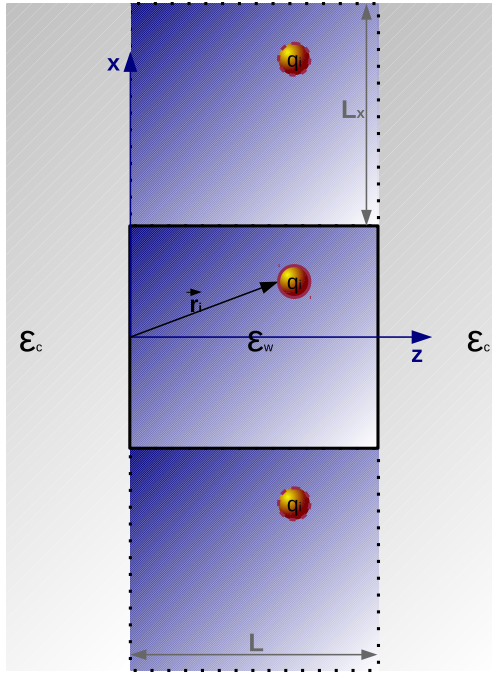}%\vspace{0.2cm}\hspace{0.1cm}
\end{center}
\caption{Representation of the system. Only the first two images of the main simulation box in $\hat{x}$ direction are shown.}
\label{fig1}
\end{figure}

Consider a point particle of charge $q_i$ at position ${\bf r}_i=(x_i,y_i,z_i)$ inside a simulation box with sides of lengths $L_x$, $L_y$, and $L$; in $x$, $y$, and $z$ directions, respectively. This system is replicated along the $x$ and $y$ axis, generating an infinite periodic charged system  of finite width $L$ in the $z$ direction. The dielectric constant in the region $0<z<L$ is $\epsilon_w$, while in the regions $z<0$ and $z>L$ it is $\epsilon_c$, see Fig.~\ref{fig1}.
%%%%%%%%%%%%%%%% figure 1 %%%%%%%%%%%%%%%%%%%%%
%%%%%%%%%%%%% end of figure %%%%%%%%%%%%%%%%%
The electrostatic potential at position ${\bf r}=(x,y,z)$ satisfies the Poisson equation
%%%%eq
\begin{equation}\label{main}
\nabla^2 G({\bf r},{\bf r}_i)=-\frac{4\pi q_i}{\epsilon_w}\sum_{m_x, m_y=-\infty}^{\infty}\delta({\pmb r}-{\pmb r}_i+m_x L_x \hat{\pmb x}+m_y L_y\hat{\pmb y})
\ ,
\end{equation}
%%%%eq
The periodic delta function can be expressed using
Fourier transform representation as
%%%%eq
\begin{eqnarray}
\sum_{m_x,m_y=-\infty}^{\infty}\delta(x-x_i+m_x L_x)\delta(y-y_i+m_y L_y)=   \nonumber \\
\frac{1}{L_x L_y}\sum_{{\pmb m}=-\infty}^{\infty}e^{i \left[ \frac{2\pi m_x}{L_x}(x-x_i)+ \frac{2\pi m_y}{L_y}(y-y_i)\right]} \ ,
\end{eqnarray}
%%%%eq
where ${\pmb m}=(m_x,m_y)$. We now write the Green function as
%%%%eq
\begin{equation}\label{Gr}
\begin{split}
&G({\bf r},{\bf r}_i)= \\
&\frac{1}{L_xL_y}\sum_{{\pmb m}=-\infty}^{\infty}g_{\pmb m}(z_i,z)e^{i \left[ \frac{2\pi m_x}{L_x}(x-x_i)+ \frac{2\pi m_y}{L_y}(y-y_i)\right]} \ ,
\end{split}
\end{equation}
%%%%eq
which is periodic in $\hat{\pmb x}$ and $\hat{\pmb y}$ directions. Inserting Eq.~\ref{Gr} into  Eq.~\ref{main} we obtain
%%%%eq
\begin{equation}
\label{green1}
\begin{split}
&\dfrac{\partial^2 g_{\pmb m}(z_i,z)}{\partial z^2}-k^2 g_{\pmb m}(z_i,z) =-\frac{4\pi q_i}{\epsilon_w}\delta(z-z_i) \ ,
\end{split}
\end{equation}
%%%%eq
where $k=2\pi\sqrt{m_x^2/L_x^2+m_y^2/L_y^2}$. The general solution of Eq.~\ref{green1} has the form $Ae^{-kz}+Be^{kz}$. The electrostatic potential must vanish as $z \rightarrow \pm \infty$, restricting  its form  in the outer regions, $z<0$ and $z>L$, to a decaying exponential. Using the symmetry properties of the Green function and the boundary conditions we obtain
%%%%eq
\begin{equation}
\begin{split}
&g_{\pmb m}(z_i,z) = \frac{2\pi q_i}{\epsilon_w k (1-\gamma^2 e^{-2 k L})}\times \\
&\left[e^{-k |z-z_i|}+ \gamma e^{-k (z+z_i)}+\gamma e^{-2 k L}e^{k (z+z_i)}+\gamma^2 e^{-2 k L}e^{k |z-z_i|}\right] \ ,
\end{split}
\end{equation}
where $\gamma=(\epsilon_w-\epsilon_c)/(\epsilon_w+\epsilon_c)$.
%%%%eq
The periodic Green function assumes the form
\begin{equation}
\label{potpure}
\begin{split}
&G({\bf r},{\bf r}_i)=\frac{1}{L_xL_y}\sum_{{\pmb m}}g_{\pmb m}(z_i,z) \times \\
&\cos \left[ \frac{2\pi m_x}{L_x}(x-x_i)+ \frac{2\pi m_y}{L_y}(y-y_i)\right] \ .
\end{split}
\end{equation}
%%%%eq

%We point out that the Eq.~\ref{potpure} diverges in the limit $k\rightarrow 0$ or $m_x$,$m_y \rightarrow 0$. To handle this term we write an auxiliary function
In the absence of dielectric contrast, $\gamma \rightarrow 0$, Eq.~\ref{potpure} reduces to
%%%%eq
\begin{equation}
\label{eqaux}
\begin{split}
&G_0({\bf r},{\bf r}_i)=\frac{2\pi q_i}{\epsilon_w L_xL_y}\sum_{{\pmb m}=-\infty}^{\infty}\frac{e^{-k |z-z_i|}}{k} \times \\
&\cos \left[ \frac{2\pi m_x}{L_x}(x-x_i)+ \frac{2\pi m_y}{L_y}(y-y_i)\right] \ , 
\end{split}
\end{equation}
%%%%eq
which is a representation of the electrostatic potential produced by a periodically replicated point charge in the $x$ and $y$ directions. Eq.~\ref{eqaux} diverges in the limit $k\rightarrow 0$, when $m_x$,$m_y \rightarrow 0$.  Although this divergence can be renormalized, the remaining sum is still slowly convergent.  We note, however, that the electrostatic potential described by Eq.~\ref{eqaux} can be efficiently calculated using a modified 3d Ewald summation technique~\cite{YeBe99,DoGi16} or
other other methods~\cite{Lek91}. The details of this calculation are presented in the appendix. With the aid of Eq.~\ref{eqaux} we can rewrite the total electrostatic potential as 
%%%%eq
\begin{equation}
G({\bf r},{\bf r}_i)=[G({\bf r},{\bf r}_i)-G_0({\bf r},{\bf r}_i)]
+G_0({\bf r},{\bf r}_i) \ .
\end{equation}
%%%%eq
We define $\tilde G({\bf r},{\bf r}_i)=G({\bf r},{\bf r}_i)-G_0({\bf r},{\bf r}_i)$ as the polarization contribution to the total Green function given by
%%%%eq
\begin{equation}
\begin{split}
&\tilde G({\bf r},{\bf r}_i)=\frac{2\pi q_i}{\epsilon_w L_xL_y}\sum_{ {\pmb m}=-\infty}^{\infty}\frac{1}{k (1-\gamma^2 e^{-2 k L})}\times \\
&\left[\gamma e^{-k (z+z_i)}+\gamma e^{-2 k L}e^{k (z+z_i)}+2\gamma^2 e^{-2 k L}\cosh{(k(z-z_i))}\right]\times \\
&\cos \left[ \frac{2\pi m_x}{L_x}(x-x_i)+ \frac{2\pi m_y}{L_y}(y-y_i)\right] \ .
\end{split}
\end{equation}
%%%%eq
The limit $k \rightarrow 0$, $m_x=m_y=0$, requires additional care.  For $-1<\gamma<1$ we find that the $m_x=m_y=0$ diverges as  
%%%%eq
\begin{equation}
\begin{split}
-\frac{4\pi q_i}{\epsilon_w L_xL_y}[\frac{\gamma}{k(\gamma-1)}+\frac{\gamma L}{(\gamma-1)^2} + \mathcal{O}(k)]  \ .
\end{split}
\end{equation}
%%%%eq
Since this is a constant, it will not contribute to the force and can be renormalized away.
%Therefore, the terms that scales with $k^n$, $n\geq 1$, vanishes and we are left with an infinity that can be renormalized away with a potential redefinition, resulting 
%%%%eq
%\begin{equation}
%\begin{split}
%G_{(\bar\gamma)}({\bf r},{\bf r}_i)=-\frac{4\pi q_i\gamma L}{\epsilon_w L_xL_y(1-\gamma)^2} \ ,
%\end{split}
%\end{equation}
%%%%eq
%that does not depend on ${\pmb r}_i$ or ${\pmb r}$, thus is a constant contribution to the potential. Consequently, we can set $G_{(\bar\gamma)}({\bf r},{\bf r}_i)=0$. 
For $\gamma=-1$, we find that $m_x=m_y=0$ term contains an infinite constant and a finite function of $z$,
%%%%eq
\begin{equation}
\begin{split}
\frac{2\pi q_i}{\epsilon_w L_xL_y}\left[-\frac{1}{k}+(z+z_i-2 \frac{z_i z}{L})+\mathcal{O}(k)\right] \ .
\end{split}
\end{equation}
%%%%eq
Once again neglecting the infinite constant, we write
%%%%eq
\begin{equation}
\begin{split}
G_{(-1)}({\bf r},{\bf r}_i)=\frac{2\pi q_i}{\epsilon_w L_xL_y}(z+z_i-2\frac{z_i z}{L}) \ .
\end{split}
\end{equation}
%%%%eq
For $\gamma= 1$  we find
%%%%eq
\begin{equation}
\begin{split}
\frac{2\pi q_i}{\epsilon_w L_xL_y}\left[\frac{2}{L k^2}-\frac{1}{k}+\frac{2 L^2-3L(z+z_i)+3(z^2+z_i^2)}{3L}+\mathcal{O}(k)\right] \ ,
\end{split}
\end{equation}
%%%%eq
so that 
%%%%eq
\begin{equation}
\begin{split}
G_{(+1)}({\bf r},{\bf r}_i)=\frac{2\pi q_i}{\epsilon_w L_xL_y}\left[-(z+z_i)+\frac{z^2+z_i^2}{L}\right] \ .
\end{split}
\end{equation}
%%%%eq
The final expression for the total electrostatic potential can now be written as
%%%%eq
\begin{equation}
\begin{split}
&G({\bf r},{\bf r}_i)= G_0({\bf r},{\bf r}_i) + G_{(\gamma)}({\bf r},{\bf r}_i)+\\ 
&\frac{2\pi q_i}{\epsilon_w L_xL_y}\sum_{{\pmb m'}=-\infty}^{\infty}\frac{1}{k (1-\gamma^2 e^{-2 k L})}\times \\
& \left(\gamma e^{-k (z+z_i)}+\gamma e^{-2 k L}e^{k (z+z_i)}+2\gamma^2 e^{-2 k L}\cosh{(k|z-z_i|)}\right)\times \\
&\cos\left[2\pi (\frac{m_x}{L_x}(x-x_i) + \frac{m_y}{L_y}(y-y_i))\right]  \ ,
\end{split}
\label{en0}
\end{equation}
%%%%eq
where the function $G_{(\gamma)}({\bf r},{\bf r}_i)$ is non zero only for $\gamma=+1$ and $-1$ and the 
prime excludes $m_x=m_y=0$ term in the summation.

The total energy for a system of $N$ periodically replicated charged particles is then given by
%%%%eq
\begin{equation}
\begin{split}
U=\sum_{i=1}^{N}\sum_{j=1}^{N} q_j\frac{G({\bf r}_j,{\bf r}_i)}{2} \ .
\end{split}
\label{en}
\end{equation}
%%%%eq
We can split the total energy into the polarization and direct Coulomb contributions 
%%%%eq
\begin{equation}
\begin{split}
U = U_{Ew} + U_{p}  \ ,
\end{split}
\end{equation}
%%%%eq
where $U_{Ew}$ is the direct Coulomb contribution,
%%%%eq
\begin{equation}
\begin{split}
U_{Ew}=\sum_{i=1}^{N}\sum_{j=1}^{N} q_j\frac{G_0({\bf r}_j,{\bf r}_i)}{2} \ ,
\end{split}
\label{ewald}
\end{equation}
%%%%eq
which can be calculated using the modified 3d Ewald summation method, see appendix.  
The energy $U_{p}$ due to surface  polarizability can be rewritten as
% To efficiently compute it, we first rescale the the dimensions of length by the Bjerrum length, defined as $\lambda_B=q^2/\epsilon_w k_BT=q^2\beta/\epsilon_w$, the energies by the thermal energy $\beta^{-1}$ and the charges by $q$, the proton charge. Furthermore, without loss of generality, we set $L_x=L_y=L_d$ and the energy takes the appropriate form of
%%%%eq
\begin{equation}
\begin{split}
&U_p= U_{\gamma}+\frac{\pi}{\epsilon_w L_d^2}\sum_{{\pmb m}'} \frac{\gamma}{k (1-\gamma^2 e^{-2 k L})}\{ f_1({\pmb m})^2 + f_2({\pmb m})^2 + \\
& +e^{-2 k L} \left(f_3({\pmb m})^2 +f_4({\pmb m})^2 \right) + 2\gamma e^{-2 k L}[f_3({\pmb m})f_1({\pmb m}) + \\
& +f_2({\pmb m})f_4({\pmb m})]\} \ ,
\end{split}
\end{equation}
%%%%eq
where without loss of generality we have set $L_x=L_y=L_d$. The number of integers, $(m_x,m_y)$, necessary to obtain a 
converged energy will depend on the lateral size of the simulation box, $L_d$. 
The contribution $U_{\gamma}$ arises from the $k \rightarrow 0$ limit, and is zero if $\gamma \neq (-1,+1)$.  For  $\gamma=-1$ we find
%
%
% \begin{equation}
% \begin{split}
% \beta U_{(\bar{\gamma})}= -\frac{2\pi\lambda_B Q_t^2}{L_d^2}[\frac{\gamma L}{(\gamma-1)^2}] \ ,
% \end{split}
% \label{eq22}
% \end{equation}
% where $Q_t=\sum_{i=1}^N q_i$. The next term is
%
%
%%%%eq
\begin{equation}
\begin{split}
U_{(-1)}=-\frac{2\pi}{L_d^2} \left[ \frac{M_z^2}{L} - Q_t M_z \right] \ ,
\end{split}
\end{equation}
%%%%eq
where $Q_t=\sum_{i=1}^N q_i$ and $M_z=\sum_{i=1}^N q_i z_i$. For $\gamma=+1$ we obtain
%%%%eq
\begin{equation}
\begin{split}
U_{(+1)}=-\frac{2\pi Q_t}{L_d^2}\left[M_z - \frac{\Omega_z}{L}\right] \ ,
\end{split}
\end{equation}
%%%%eq
where $\Omega_z=\sum_{i=1}^Nq_i z_i^2$. The $f_i({\pmb m})$ 
functions are defined as 
%%%%eq
\begin{equation}
f_1({\pmb m})=\sum_{i=1}^{N} q_i \cos\left[\frac{2\pi}{L_d}(m_xx_i+m_yy_i)\right]e^{-kz_i} \ ,
\end{equation}
%%%%eq
%%%%eq
\begin{equation}
f_2({\pmb m})=\sum_{i=1}^{N} q_i \sin\left[\frac{2\pi}{L_d}(m_xx_i+m_yy_i)\right]e^{-kz_i} \ ,
\end{equation}
%%%%eq
%%%%eq
\begin{equation}
f_3({\pmb m})=\sum_{i=1}^{N} q_i \cos\left[\frac{2\pi}{L_d}(m_xx_i+m_yy_i)\right]e^{kz_i} \ ,
\end{equation}
%%%%eq
%%%%eq
\begin{equation}
f_4({\pmb m})=\sum_{i=1}^{N} q_i \sin\left[\frac{2\pi}{L_d}(m_xx_i+m_yy_i)\right]e^{kz_i} \ .
\end{equation}
%%%%eq
Note that $k$ depends on ${\pmb m}$ and the $f$ functions must be updated for each particle move. 
There is, however, no need to recalculate all the functions, but only the contribution to each function that depends on the position of the particle that is being moved.  This makes the energy update very efficient. Furthermore, the prefactors that depend on the exponential functions of $m_x$ and $m_y$ can be precalculated at the beginning of the simulation. Finally, if there is a surface charge present at the interfaces, it can be included as an external potential, see the appendix and Ref.~\cite{DoGi16},  
%%%%eq
\begin{equation}\label{ersur}
\begin{split}
 U_{sur}=-\frac{2\pi(\sigma_1-\sigma_2)}{\epsilon_w}\sum_{i=1}^N q_i z_i \ ,
\end{split}
\end{equation}
%%%%eq
where $\sigma_1$ and $\sigma_2$ are the surface charge densities at $z=0$ and $z=L$, respectively.
%%%%%%%%%%%%%%%% figure 1 %%%%%%%%%%%%%%%%%%%%%
\begin{figure}[h]
\vspace{0.5cm}
\begin{center}
\includegraphics[scale=0.35]{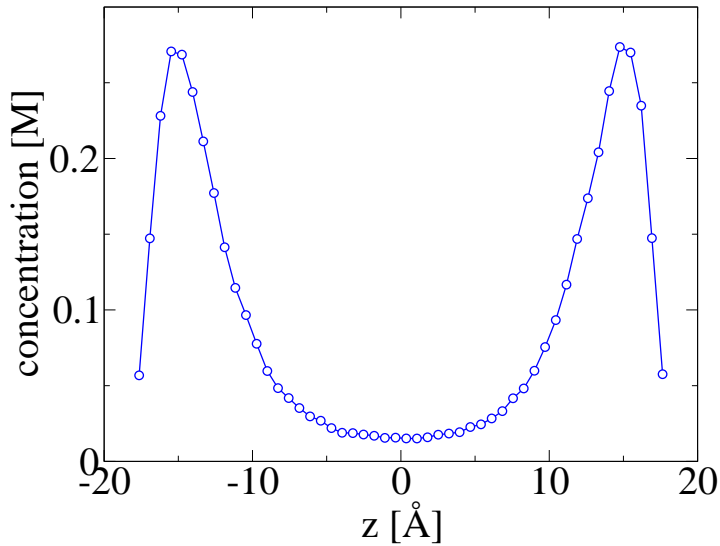}%\vspace{0.2cm}\hspace{0.1cm}
\end{center}
\caption{Density profile of trivalent counterions confined between charged dielectric surfaces, $\gamma=0.95$. The surfaces charge densities are $-0.05~$C$/$m$^2$. The line is a guide to the eyes.}
\label{fig2}
\end{figure}
%%%%%%%%%%%%% end of figure %%%%%%%%%%%%%%%%%
% where $U_{pp}$ is the electrostatic energy of the charged plates related with the polarization and $U_{up}$ without this feature.
% \begin{equation}
% \begin{split}
% &\beta U_{up}=-2\pi\lambda_B(\sigma_1-\sigma_2)\sum_{i=1}^N q_i z_i-\\
% &2\pi\lambda_B Q_t\frac{(\sigma_1+\sigma_2)}{2} L - 2\pi\lambda_B \sigma_1\sigma_2L_d^2 L \ ,
% \end{split}
% \end{equation}
% where $\sigma_1$ and $\sigma_2$ are the charge densities of the surfaces located at $z=0$ and $z=L$, respectively. Considering the sum of infinity images of the charged surfaces
% we can derive the energy $U_{pp}$ which can be written as
% \begin{equation}
% \begin{split}
% &\beta U_{pp}=- 4\pi\lambda_B Q_t (\sigma_1+\sigma_2)\frac{\gamma L}{(\gamma-1)^2}-\\
% &2\pi\lambda_B L_d^2 (\sigma_1+\sigma_2)^2\frac{\gamma L}{(\gamma-1)^2} \ .
% \end{split}
% \label{eq2}
% \end{equation}
% Considering that $Q_t=-(\sigma_1+\sigma_2)L_d^2$, the terms in Eq.~\ref{eq2} and \ref{eq22} cancel each other for the interval of $\gamma=(-1,+1)$. In the interval $\gamma=[+1]$ this terms diverges. In the interval $\gamma=[-1]$ this term is $\beta U_{pp} = - \pi\lambda_B L_d^2 (\sigma_1+\sigma_2)^2 L /2$

\section{Simulations and Results}

To demonstrate the utility of the new simulation method,
we perform Monte Carlo simulations of an electrolyte solution in the $NVT$ ensemble using Metropolis algorithm~\cite{NiTe53}. To efficiently sample the phase space we use both short and long displacement moves~\cite{AlTi87,Fr02}. The effective ionic radii are set to $r_c=2~$\AA. The Bjerrum length, defined as $q^2\beta/\epsilon_w$, where $\beta$ is the inverse thermal energy and $q$ is the proton charge, is set to $7.2~$\AA, typical value for water at room temperature. The system relaxes to equilibrium in $1\times10^6$ Monte Carlo steps. The ionic density profiles are obtained using $1\times10^5$ uncorrelated samples.

In Fig.~\ref{fig2} we show the density profile of trivalent counterions confined between charged dielectric surfaces of $\gamma=0.95$. The confining surfaces are separated by a distance $L=40~$\AA. The number of counterions is $N_c=100$ and the surfaces are equally charged with charge density $-0.05~$C$/$m$^2$. We see a strong repulsion of ions from the interface produced by the induced surface charge. This result is in agreement with an earlier image charge algorithm~\cite{DoLe15}. However, the present method is an order of magnitude more efficient.

\begin{figure}[h]
\vspace{0.5cm}
\begin{center}
\includegraphics[scale=0.35]{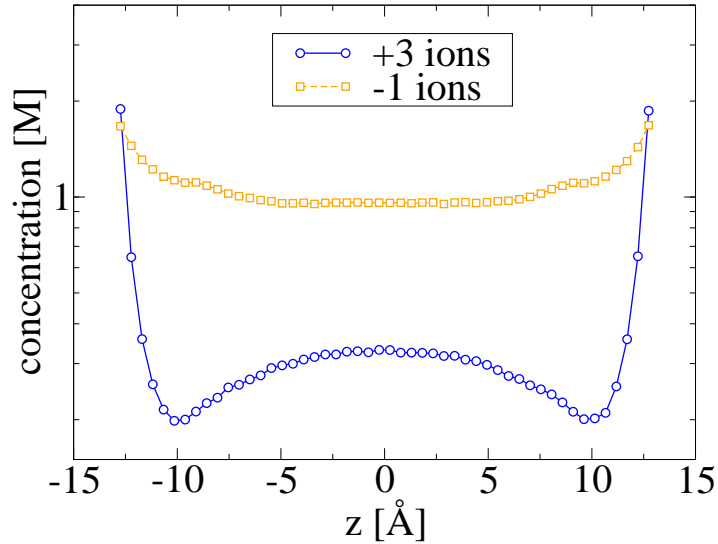}%\vspace{0.5cm}%\hspace{0.1cm}
\end{center}
\caption{Density profiles of cations and anions confined between grounded metal surfaces, $\gamma=-1$. The $3:1$ salt concentration is $0.35~$M. The lines are guides to the eye.}
\label{fig3}
\end{figure}

In Fig.~\ref{fig3} we show the density profiles of cations and anions of a dissolved 3:1 electrolyte  at concentration $0.35~$M, confined by grounded metal electrodes, $\gamma=-1$, separated by distance $L=30~$\AA.
%%%%%%%%%%%%%%%% figure 1 %%%%%%%%%%%%%%%%%%%%%
%%%%%%%%%%%%% end of figure %%%%%%%%%%%%%%%%%
Now, instead of the repulsion of the previous case, we see the expected attraction of charges to the metal electrodes. This effect can be understood considering the image charges of opposite sign induced inside the electrodes. 

\begin{figure}[h]
\vspace{0.5cm}
\begin{center}
\includegraphics[scale=0.35]{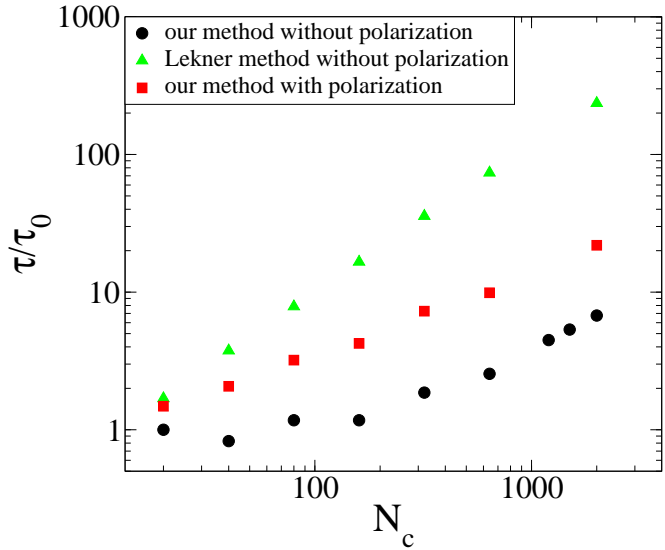}%\vspace{0.5cm}%\hspace{0.1cm}
\end{center}
\caption{CPU time to perform $10^6$ energy updates as a function of the number of particles in the system. The distance between the polarizable plates is $L=10$\AA, with $\gamma=0.95$. The Bjerrum length was set to $\lambda_B=14.5$\AA, the superficial charge to $\sigma=-0.12$C$/$m$^2$ and ionic radius to $2$\AA. }
\label{fig4}
\end{figure}

Finally, in Fig.~\ref{fig4} we compare the characteristic CPU times of our simulation method with a standard implementation of Lekner summation which does not account for polarization~\cite{Lek91}. We see that for reasonably large system sizes,  Lekner summation is at least an order of magnitude slower than our method. Furthermore, for large $N_c$ we see that even for systems with polarization our
method remains an order of magnitude faster than Lekner summation without polarization.
%%%%%%%%%%%%%%%% figure 1 %%%%%%%%%%%%%%%%%%%%%
%%%%%%%%%%%%% end of figure %%%%%%%%%%%%%%%%% 

\section{Conclusions}

We have presented an efficient new method for simulating Coulomb systems confined by polarizable surfaces. The method relies on the exact solution of Poisson equation in terms of periodic Green functions. We were able to separate the electrostatic energy into polarization and direct Coulomb contributions.  The latter can be efficiently calculated using a modified Ewald method developed in the previous work~\cite{DoGi16}.  The polarization energy is separated into terms which can be locally updated for each particle move without the need of recalculating the whole electrostatic energy. 
The results of the new simulation method were compared with the earlier approach~\cite{DoLe15} 
and found to lead to identical
ionic density profiles, with a significant gain in simulation time. The advantage of the new
method is that it is very fast and easy to implement with a simple adaptation of the usual 3d Ewald summation code for either Monte Carlo or Molecular Dynamics simulations. Finally, we note that the calculations presented in this paper can be easily extended to study systems with two confining walls of 
distinct dielectric constants.

\section{Acknowledgments}

This work was partially supported by the CNPq, CAPES, Alexander von Humboldt Foundation, INCT-FCx, and by the US-AFOSR under the grant FA9550-16-1-0280.

\appendix

\section{Ewald summation in slab geometry}

For systems in slab geometry, without dielectric discontinuities, there are well established algorithms~\cite{Kl92,KaMi01,YeBe99,Sp97,Lek89,Spe94,ArHo02,PiMa02}. Recently, we developed an efficient algorithm~\cite{DoGi16} where the surface charge at the slab boundaries is treated as an external potential, speeding up the traditional simulations in which the surface charges are modeled by point particles. We briefly discuss how this modified Ewald method can be used to calculate the electrostatic potential produced by a periodically replicated point charge.  We start by considering an isotropic system replicated in all three dimensions and then take the slab geometry limit, in which one of the directions grows much slower than the other two. Consider $N$ particles of charge q$^j$ confined in a cell of lengths $L_x$, $L_y$ and $L_z$. The infinite system is constructed with the definition of the replication vector ${\pmb r}_{rep}=(n_xL_x,n_yL_y,n_zL_z)$, where $n$'s span the positive and negative integers. The electrostatic potential produced by the ions and all the replicas at position ${\pmb r}$ can be written as 
%%%%eq
\begin{eqnarray}\label{elec_pot}
\phi({\pmb r})=\sum_{\pmb n}^{\infty}\sum_{j=1}^{N}\int\frac{\rho^j({\pmb s})}{\epsilon_w |{\pmb r}-{\pmb s}|}d^3{\pmb s} \ ,
\end{eqnarray}
%%%%eq
where $\rho^j({\pmb s})=\text{q}^j \delta({\pmb s}-{\pmb r}^j-{\pmb r}_{ep})$ is the charge density of q$^j$ and its replicas. Adding and subtracting a Gaussian charge density distribution on top of each charge q$^j$ we can split the potential into long and short range contributions, writing:
%%%%eq
\begin{eqnarray}\label{elec_pot_E}
\phi({\pmb r})=\sum_{\pmb n}^{\infty}\sum_{j=1}^{N}\int\frac{\rho^j_G({\pmb s})}{\epsilon_w |{\pmb r}-{\pmb s}|}d^3{\pmb s} + \nonumber \\
\sum_{\pmb n}^{\infty}\sum_{j=1}^{N}\int\frac{\rho^j({\pmb s})-\rho^j_G({\pmb s})}{\epsilon_w |{\pmb r}-{\pmb s}|}d^3{\pmb s}  \ ,
\end{eqnarray}
%%%%eq
where $\rho^j_G({\pmb s})=\text{q}^j (\kappa_e^3/\sqrt{\pi^3})\exp{(-\kappa_e^2|{\pmb s}-{\pmb r}^j-{\pmb r}_{rep}|^2)}$ and $\kappa_e$ is a damping parameter. The first term on the right hand side of Eq.~\ref{elec_pot_E} is long ranged (it has a non integrable tail) and can be efficiently summed using Fourier representation. The second term can be rewritten using the Complementary Error Function. The electrostatic potential then takes the form
%%%%eq
\begin{equation}\label{elec_pot_E_f}
\begin{split}
&\phi({\pmb r})=\sum_{{\pmb k}={\pmb 0}}^{\infty}\sum_{j=1}^{N}\frac{4\pi \text{q}^j}{\epsilon_w V |{\pmb k}|^2}\exp{[-\frac{|{\pmb k}|^2}{4\kappa_e^2}+i{\pmb k}\cdot({\pmb r}-{\pmb r}^j)]} + \\
&\sum_{j=1}^{N}\text{q}^j\frac{\text{erfc}(\kappa_e|{\pmb r}-{\pmb r}^j|)}{\epsilon_w |{\pmb r}-{\pmb r}^j|} \ ,
\end{split}
\end{equation}
%%%%eq
where ${\pmb k}=(\frac{2\pi}{L_x}n_1,\frac{2\pi}{L_y}n_2,\frac{2\pi}{L_z}n_3)$ and $V=L_xL_yL_z$, the volume of the main cell.  Since the second term is short ranged it can be treated using simple periodic boundary conditions, as long as $\kappa_e$ is sufficiently large.

The first term of the Fourier series diverges when ${\pmb k}\rightarrow 0$. To understand better the significance of this divergence  we study this term separately by expanding it around the $k=0$. We write
 %%%%eq
\begin{equation}\label{pot}
\begin{split}
&\lim_{{\pmb k} \rightarrow 0}\sum_{j=1}^{N}\text{q}^j\frac{1}{|{\pmb k}|^2}-\sum_{j=1}^{N}\text{q}^j\frac{1}{4\kappa_e^2}+
\lim_{{\pmb k} \rightarrow 0}\sum_{j=1}^{N}\text{q}^j\frac{i{\pmb k}\cdot({\pmb r}-{\pmb r}^j)}{|{\pmb k}|^2} \\
& - \lim_{{\pmb k} \rightarrow 0}\sum_{j=1}^{N}\text{q}^j\dfrac{[{\pmb k}\cdot({\pmb r}-{\pmb r}^j)]^2}{2|{\pmb k}|^2} + \mathcal{O}(|{\pmb k}|) \ .
\end{split}
\end{equation}
%%%%eq
If the system is non neutral, it is possible to renormalize the two diverging constant terms by redefining the zero of the electrostatic potential. Consequently, we can neglect the infinite constants which do not influence the physics of the system. However, the third and fourth term have dependence on particle positions and hence must be properly accounted for. The third sum on the right can be written as 
%%%%eq
\begin{eqnarray}\label{s3}
S_3=\sum_{j=1}^{N}\text{q}^j\int_{-\infty}^{+\infty} \delta({\pmb k})\frac{i{\pmb k}\cdot({\pmb r}-{\pmb r}^j)}{|{\pmb k}|^2}d{\pmb k} \ ,
\end{eqnarray}
%%%%eq 
where we use the delta representation $\delta({\pmb k})=(2\pi)^{-3}\int_{-{\pmb H}}^{{\pmb H}} e^{i{\pmb k}\cdot{\pmb p}}d^3p$. The limits in delta integration, $-{\bf H}$ to ${\bf H}$, where ${\bf H}=(H_1,H_2,H_3)$, must be performed in accordance with the real space sum. We define $H_1=\alpha_1 L_c$, $H_2=\alpha_2 L_c$ and $H_3=\alpha_3 L_c$, where $L_c$ is some characteristic macroscopic length scale. For an isotropic bulk systems  $H$'s grow at the same rate. On the other hand, for systems with a slab geometry $H_1$ and $H_2$ should grow much faster than $H_3$. Explicitly performing the integrals over $p$'s we obtain
%%%%eq
\begin{equation}
\delta({\pmb k})=\frac{1}{(2 \pi)^3}\prod_{i=1}^3\int_{-\alpha_i\frac{L_c}{2}}^{\alpha_i\frac{L_c}{2}} e^{i k_i p_i} dp_i=\frac{1}{ \pi^3}\prod_{i=1}^3\frac{\text{sin}(k_i\alpha_i L_c/2)}{k_i} \ ,
\end{equation}
%%%%eq
and Eq.~\ref{s3} can now be written as $S_3=\sum_{j=1}^{N} q_j {\pmb D} \cdot ({\pmb r}-{\pmb r}^j)$, where the components of the vector ${\pmb D}$ are
%%%%eq
\begin{eqnarray}
D_n= \frac{i}{\pi^3}\int_{-\infty}^{+\infty} \frac{ k_n}{|{\pmb k}|^2} \prod_{j=1}^3\frac{\text{sin}(k_j\alpha_j L_c/2)}{k_j}d^3{\pmb k}\ ,
\end{eqnarray}
%%%%eq
which by symmetry integrates to  zero, $D_n=0$, so that $S_3=0$. The last term can be written as
%%%%eq
\begin{equation}
S_4=-\sum_{j=1}^{N}\text{q}^j\int_{-\infty}^{+\infty}\delta({\pmb k})\dfrac{[{\pmb k}\cdot({\pmb r}-{\pmb r}^j)]^2}{2|{\pmb k}|^2}d^3{\pmb k} \ .
\end{equation}    
%%%%eq
Applying once again the delta function representation, we obtain
%%%%eq
\begin{equation}
S_4=-\sum_{j=1}^{N}\frac{\text{q}^j}{2  \pi^3}\sum_{n=1}^3 B_n(r_n-r^j_n)^2 \ ,
\end{equation}
%%%%eq
where the index $n$ corresponds to the $x$, $y$, and $z$ components of the vector ${\pmb r}$ and
%%%%eq
\begin{equation}
B_n= \int_{-\infty}^{+\infty} d^3{\pmb k} \frac{ k_n^2}{|{\pmb k}|^2} \prod_{j=1}^3\frac{\text{sin}(k_j\alpha_j L_c/2)}{k_j}\ .
\end{equation}
%%%%eq
The coefficients $B_n$ can be simplified to~\cite{Sm81}
%%%%eq
\begin{equation}
B_1=\dfrac{\pi^{\frac{5}{2}}}{2}\int_{0}^{+\infty} \dfrac{\alpha_{13} e^{-\frac{\alpha_{13}^2}{4 t}}\text{erf}(\frac{\alpha_{23}}{2\sqrt{t}})\text{erf}(\frac{1}{2\sqrt{t}})}{t^{\frac{3}{2}}} dt \ ,
\end{equation}
%%%%eq
%%%%eq
\begin{equation}
B_2=\dfrac{\pi^{\frac{5}{2}}}{2}\int_{0}^{+\infty} \dfrac{\alpha_{23} e^{-\frac{\alpha_{23}^2}{4 t}}\text{erf}(\frac{\alpha_{13}}{2\sqrt{t}})\text{erf}(\frac{1}{2\sqrt{t}})}{t^{\frac{3}{2}}} dt \ ,
\end{equation}
%%%%eq
%%%%eq
\begin{equation}
B_3=\dfrac{\pi^{\frac{5}{2}}}{2}\int_{0}^{+\infty} \dfrac{e^{-\frac{1}{4 t}}\text{erf}(\frac{\alpha_{13}}{2\sqrt{t}})\text{erf}(\frac{\alpha_{23}}{2\sqrt{t}})}{t^{\frac{3}{2}}} dt \ ,
\end{equation}
%%%%eq
where $\alpha_{ij}=\alpha_i/\alpha_j$ are the aspect ratios of the macroscopic system. The coefficients $B_n$ can now be easily calculated using numerical integration. For a spherically symmetric summation of replicas the aspect ratios are $\alpha_{13}=L_x/L_z$ and $\alpha_{23}=L_y/L_z$. On the other hand, for a plane-wise summation of a slab geometry,  $\alpha_{13}\rightarrow\infty$ and $\alpha_{23}\rightarrow\infty$. In this case the integrals can be performed explicitly~\cite{Sm81}, yielding $B_1=B_2=0$ and $B_3=\pi^3$. Thus, for slab geometry we have the renormalized electrostatic potential
%%%%eq
\begin{equation}\label{elec_pot_E_new2}
\begin{split}
& \Delta \phi({\pmb r})=\sum_{{\pmb k}\neq{\pmb 0}}^{\infty}\sum_{j=1}^{N}\frac{4\pi \text{q}^j}{\epsilon_w V |{\pmb k}|^2}\exp{[-\frac{|{\pmb k}|^2}{4\kappa_e^2}+i{\pmb k}\cdot({\pmb r}-{\pmb r}^j)]} \\
& -\sum_{j=1}^{N}\frac{2\pi \text{q}^j}{\epsilon_w V}(r_3-r_3^j)^2 + 
\sum_{j=1}^{N}\text{q}^j\frac{\text{erfc}(\kappa_e|{\pmb r}-{\pmb r}^j|)}{\epsilon_w |{\pmb r}-{\pmb r}^j|} \ , 
\end{split}
\end{equation}
%%%%eq
and the energy, $U_{Ew}=\dfrac{1}{2}\sum_{i=1}^N\text{q}^i\Delta \phi({\pmb r}^ i)$, is  
%%%%eq
\begin{equation}\label{ener}
\begin{split}
U_{Ew}=\sum_{{\pmb k}\neq{\pmb 0}}^{\infty}\frac{2\pi}{\epsilon_w V |{\pmb k}|^2}\exp{[-\frac{|{\pmb k}|^2}{4\kappa_e^2}]}[A({\pmb k})^2+B({\pmb k})^2]\\ 
+ \frac{2\pi}{\epsilon_w V}[M_z^2-Q_t\Omega_z] \\
+ 
\dfrac{1}{2}\sum_{i \ne j}^N\text{q}^i\text{q}^j\frac{\text{erfc}(\kappa_e|{\pmb r}^i-{\pmb r}^j|)}{\epsilon_w |{\pmb r}^i-{\pmb r}^j|} - \frac{\kappa_e}{\epsilon_w \sqrt{\pi}}\sum_i^Nq_i^2\ , 
\end{split}
\end{equation}
%%%%eq
where
%%%%eq
\begin{eqnarray}
A({\pmb k})=\sum_{i=1}^N \text{q}^i\text{cos}({\pmb k}\cdot{\pmb r}^i) \ , \nonumber \\
B({\pmb k})=-\sum_{i=1}^N \text{q}^i\text{sin}({\pmb k}\cdot{\pmb r}^i) \ , \nonumber \\
M_z=\sum_{i=1}^N \text{q}^i z_i \ , \nonumber \\
Q_t=\sum_{i=1}^N \text{q}^i \ , \nonumber \\
\Omega_z=\sum_{i=1}^N \text{q}^i z_i^2 \ .
\end{eqnarray}
%%%%eq
If there are surface charge densities present at the interfaces, an additional term, Eq.~\ref{ersur}, must be included. Eq.~\ref{ener} provides an efficient way of calculating the slowly converging sum in
Eq.~\ref{eqaux} allowing us to rapidly calculate the direct contribution to the total electrostatic energy, Eq.~\ref{ewald}.

\bibliography{ref.bib}

\begin{thebibliography}{65}%
\makeatletter
\providecommand \@ifxundefined [1]{%
 \@ifx{#1\undefined}
}%
\providecommand \@ifnum [1]{%
 \ifnum #1\expandafter \@firstoftwo
 \else \expandafter \@secondoftwo
 \fi
}%
\providecommand \@ifx [1]{%
 \ifx #1\expandafter \@firstoftwo
 \else \expandafter \@secondoftwo
 \fi
}%
\providecommand \natexlab [1]{#1}%
\providecommand \enquote  [1]{``#1''}%
\providecommand \bibnamefont  [1]{#1}%
\providecommand \bibfnamefont [1]{#1}%
\providecommand \citenamefont [1]{#1}%
\providecommand \href@noop [0]{\@secondoftwo}%
\providecommand \href [0]{\begingroup \@sanitize@url \@href}%
\providecommand \@href[1]{\@@startlink{#1}\@@href}%
\providecommand \@@href[1]{\endgroup#1\@@endlink}%
\providecommand \@sanitize@url [0]{\catcode `\\12\catcode `\$12\catcode
  `\&12\catcode `\#12\catcode `\^12\catcode `\_12\catcode `\%12\relax}%
\providecommand \@@startlink[1]{}%
\providecommand \@@endlink[0]{}%
\providecommand \url  [0]{\begingroup\@sanitize@url \@url }%
\providecommand \@url [1]{\endgroup\@href {#1}{\urlprefix }}%
\providecommand \urlprefix  [0]{URL }%
\providecommand \Eprint [0]{\href }%
\providecommand \doibase [0]{http://dx.doi.org/}%
\providecommand \selectlanguage [0]{\@gobble}%
\providecommand \bibinfo  [0]{\@secondoftwo}%
\providecommand \bibfield  [0]{\@secondoftwo}%
\providecommand \translation [1]{[#1]}%
\providecommand \BibitemOpen [0]{}%
\providecommand \bibitemStop [0]{}%
\providecommand \bibitemNoStop [0]{.\EOS\space}%
\providecommand \EOS [0]{\spacefactor3000\relax}%
\providecommand \BibitemShut  [1]{\csname bibitem#1\endcsname}%
\let\auto@bib@innerbib\@empty
%</preamble>
\bibitem [{\citenamefont {{M. P. Allen and D. J. Tildesley}}(1987)}]{AlTi87}%
  \BibitemOpen
  \bibfield  {author} {\bibinfo {author} {\bibnamefont {{M. P. Allen and D. J.
  Tildesley}}},\ }\href@noop {} {\emph {\bibinfo {title} {Computer Simulations
  of Liquids}}}\ (\bibinfo  {publisher} {Oxford: Oxford University Press},\
  \bibinfo {address} {New York},\ \bibinfo {year} {1987})\BibitemShut {NoStop}%
\bibitem [{\citenamefont {{D. Frenkel and B. Smit}}(2002)}]{Fr02}%
  \BibitemOpen
  \bibfield  {author} {\bibinfo {author} {\bibnamefont {{D. Frenkel and B.
  Smit}}},\ }\href@noop {} {\emph {\bibinfo {title} {Understanding Molecular
  Simulation - From Algorithms to Applications}}}\ (\bibinfo  {publisher}
  {Academic Press},\ \bibinfo {address} {525 B Street, Suite 1900, San Diego,
  California 92101-4495, USA},\ \bibinfo {year} {2002})\ pp.\ \bibinfo {pages}
  {291--306}\BibitemShut {NoStop}%
\bibitem [{\citenamefont {Ewald}(1921)}]{Ewa21}%
  \BibitemOpen
  \bibfield  {author} {\bibinfo {author} {\bibfnamefont {P.}~\bibnamefont
  {Ewald}},\ }\bibfield  {title} {\enquote {\bibinfo {title} {Die berechnung
  optischer und elektrostatischer gitterpotentiale},}\ }\href@noop {}
  {\bibfield  {journal} {\bibinfo  {journal} {Ann. Phys.}\ }\textbf {\bibinfo
  {volume} {369}},\ \bibinfo {pages} {253--287} (\bibinfo {year}
  {1921})}\BibitemShut {NoStop}%
\bibitem [{\citenamefont {Essmann}\ \emph {et~al.}(1995)\citenamefont
  {Essmann}, \citenamefont {Perera}, \citenamefont {Berkowitz}, \citenamefont
  {Darden}, \citenamefont {Lee},\ and\ \citenamefont {Pedersen}}]{PeLe95}%
  \BibitemOpen
  \bibfield  {author} {\bibinfo {author} {\bibfnamefont {U.}~\bibnamefont
  {Essmann}}, \bibinfo {author} {\bibfnamefont {L.}~\bibnamefont {Perera}},
  \bibinfo {author} {\bibfnamefont {M.~L.}\ \bibnamefont {Berkowitz}}, \bibinfo
  {author} {\bibfnamefont {T.}~\bibnamefont {Darden}}, \bibinfo {author}
  {\bibfnamefont {H.}~\bibnamefont {Lee}}, \ and\ \bibinfo {author}
  {\bibfnamefont {L.~G.}\ \bibnamefont {Pedersen}},\ }\bibfield  {title}
  {\enquote {\bibinfo {title} {A smooth particle mesh ewald method},}\
  }\href@noop {} {\bibfield  {journal} {\bibinfo  {journal} {J. Chem. Phys.}\
  }\textbf {\bibinfo {volume} {103}},\ \bibinfo {pages} {8577--8593} (\bibinfo
  {year} {1995})}\BibitemShut {NoStop}%
\bibitem [{\citenamefont {Darden}, \citenamefont {York},\ and\ \citenamefont
  {Pedersen}(1993)}]{YoPe93}%
  \BibitemOpen
  \bibfield  {author} {\bibinfo {author} {\bibfnamefont {T.}~\bibnamefont
  {Darden}}, \bibinfo {author} {\bibfnamefont {D.}~\bibnamefont {York}}, \ and\
  \bibinfo {author} {\bibfnamefont {L.}~\bibnamefont {Pedersen}},\ }\bibfield
  {title} {\enquote {\bibinfo {title} {Particle mesh ewald: An $n\dot \log(n)$
  method for ewald sums in large systems},}\ }\href@noop {} {\bibfield
  {journal} {\bibinfo  {journal} {J. Chem. Phys.}\ }\textbf {\bibinfo {volume}
  {98}},\ \bibinfo {pages} {10089--10092} (\bibinfo {year} {1993})}\BibitemShut
  {NoStop}%
\bibitem [{\citenamefont {Kolafa}\ and\ \citenamefont {Perram}(1992)}]{KoPe92}%
  \BibitemOpen
  \bibfield  {author} {\bibinfo {author} {\bibfnamefont {J.}~\bibnamefont
  {Kolafa}}\ and\ \bibinfo {author} {\bibfnamefont {J.~W.}\ \bibnamefont
  {Perram}},\ }\bibfield  {title} {\enquote {\bibinfo {title} {Cutoff errors in
  the ewald summation formulae for point charge systems},}\ }\href@noop {}
  {\bibfield  {journal} {\bibinfo  {journal} {Mol. Simul.}\ }\textbf {\bibinfo
  {volume} {9}},\ \bibinfo {pages} {351--368} (\bibinfo {year}
  {1992})}\BibitemShut {NoStop}%
\bibitem [{\citenamefont {Deserno}\ and\ \citenamefont
  {Holm}(1998{\natexlab{a}})}]{DeCh98}%
  \BibitemOpen
  \bibfield  {author} {\bibinfo {author} {\bibfnamefont {M.}~\bibnamefont
  {Deserno}}\ and\ \bibinfo {author} {\bibfnamefont {C.}~\bibnamefont {Holm}},\
  }\bibfield  {title} {\enquote {\bibinfo {title} {How to mesh up ewald sums.
  i. an accurate error estimate for the particle-particle-particle-mesh
  algorithm},}\ }\href@noop {} {\bibfield  {journal} {\bibinfo  {journal} {J.
  Chem. Phys.}\ }\textbf {\bibinfo {volume} {109}},\ \bibinfo {pages}
  {7678--7693} (\bibinfo {year} {1998}{\natexlab{a}})}\BibitemShut {NoStop}%
\bibitem [{\citenamefont {Deserno}\ and\ \citenamefont
  {Holm}(1998{\natexlab{b}})}]{DeCh982}%
  \BibitemOpen
  \bibfield  {author} {\bibinfo {author} {\bibfnamefont {M.}~\bibnamefont
  {Deserno}}\ and\ \bibinfo {author} {\bibfnamefont {C.}~\bibnamefont {Holm}},\
  }\bibfield  {title} {\enquote {\bibinfo {title} {How to mesh up ewald sums.
  ii. an accurate error estimate for the particle-particle-particle-mesh
  algorithm},}\ }\href@noop {} {\bibfield  {journal} {\bibinfo  {journal} {J.
  Chem. Phys.}\ }\textbf {\bibinfo {volume} {109}},\ \bibinfo {pages}
  {7694--7701} (\bibinfo {year} {1998}{\natexlab{b}})}\BibitemShut {NoStop}%
\bibitem [{\citenamefont {Lekner}(1991)}]{Lek91}%
  \BibitemOpen
  \bibfield  {author} {\bibinfo {author} {\bibfnamefont {J.}~\bibnamefont
  {Lekner}},\ }\bibfield  {title} {\enquote {\bibinfo {title} {Summation of
  coulomb fields in computer-simulated disordered systems},}\ }\href@noop {}
  {\bibfield  {journal} {\bibinfo  {journal} {Phys. A}\ }\textbf {\bibinfo
  {volume} {176}},\ \bibinfo {pages} {485--498} (\bibinfo {year}
  {1991})}\BibitemShut {NoStop}%
\bibitem [{\citenamefont {Widmann}\ and\ \citenamefont {Adolf}(1997)}]{WiAd97}%
  \BibitemOpen
  \bibfield  {author} {\bibinfo {author} {\bibfnamefont {A.~H.}\ \bibnamefont
  {Widmann}}\ and\ \bibinfo {author} {\bibfnamefont {D.~B.}\ \bibnamefont
  {Adolf}},\ }\bibfield  {title} {\enquote {\bibinfo {title} {A comparison of
  ewald summation techniques for planar surfaces},}\ }\href@noop {} {\bibfield
  {journal} {\bibinfo  {journal} {Comput. Phys. Commun.}\ }\textbf {\bibinfo
  {volume} {107}},\ \bibinfo {pages} {167--186} (\bibinfo {year}
  {1997})}\BibitemShut {NoStop}%
\bibitem [{\citenamefont {Mazars}(2005)}]{Maz05}%
  \BibitemOpen
  \bibfield  {author} {\bibinfo {author} {\bibfnamefont {M.}~\bibnamefont
  {Mazars}},\ }\bibfield  {title} {\enquote {\bibinfo {title} {Lekner
  summations and ewald summations for quasi-two-dimensional systems},}\
  }\href@noop {} {\bibfield  {journal} {\bibinfo  {journal} {Mol. Phys.}\
  }\textbf {\bibinfo {volume} {103}},\ \bibinfo {pages} {1241--1260} (\bibinfo
  {year} {2005})}\BibitemShut {NoStop}%
\bibitem [{\citenamefont {Fedorov}\ and\ \citenamefont
  {Kornishev}(2014)}]{FeKo14}%
  \BibitemOpen
  \bibfield  {author} {\bibinfo {author} {\bibfnamefont {M.~V.}\ \bibnamefont
  {Fedorov}}\ and\ \bibinfo {author} {\bibfnamefont {A.~A.}\ \bibnamefont
  {Kornishev}},\ }\bibfield  {title} {\enquote {\bibinfo {title} {Ionic liquids
  at electrified interfaces},}\ }\href@noop {} {\bibfield  {journal} {\bibinfo
  {journal} {Chem. Rev.}\ }\textbf {\bibinfo {volume} {114}},\ \bibinfo {pages}
  {2978--3036} (\bibinfo {year} {2014})}\BibitemShut {NoStop}%
\bibitem [{\citenamefont {Reed}, \citenamefont {Lanning},\ and\ \citenamefont
  {Madden}(2007)}]{LaMa07}%
  \BibitemOpen
  \bibfield  {author} {\bibinfo {author} {\bibfnamefont {S.}~\bibnamefont
  {Reed}}, \bibinfo {author} {\bibfnamefont {O.}~\bibnamefont {Lanning}}, \
  and\ \bibinfo {author} {\bibfnamefont {P.}~\bibnamefont {Madden}},\
  }\bibfield  {title} {\enquote {\bibinfo {title} {Electrochemical interface
  between an ionic liquid and a model metallic electrode},}\ }\href@noop {}
  {\bibfield  {journal} {\bibinfo  {journal} {J. Chem. Phys.}\ }\textbf
  {\bibinfo {volume} {126}},\ \bibinfo {pages} {084704} (\bibinfo {year}
  {2007})}\BibitemShut {NoStop}%
\bibitem [{\citenamefont {Lian}\ \emph {et~al.}(2016)\citenamefont {Lian},
  \citenamefont {Liu}, \citenamefont {Aken}, \citenamefont {Gogotsi},
  \citenamefont {Wesolowski}, \citenamefont {Liu}, \citenamefont {Jiang},\ and\
  \citenamefont {Wu}}]{JiWu16}%
  \BibitemOpen
  \bibfield  {author} {\bibinfo {author} {\bibfnamefont {C.}~\bibnamefont
  {Lian}}, \bibinfo {author} {\bibfnamefont {K.}~\bibnamefont {Liu}}, \bibinfo
  {author} {\bibfnamefont {K.~L.~V.}\ \bibnamefont {Aken}}, \bibinfo {author}
  {\bibfnamefont {Y.}~\bibnamefont {Gogotsi}}, \bibinfo {author} {\bibfnamefont
  {D.~J.}\ \bibnamefont {Wesolowski}}, \bibinfo {author} {\bibfnamefont
  {H.~L.}\ \bibnamefont {Liu}}, \bibinfo {author} {\bibfnamefont {D.~E.}\
  \bibnamefont {Jiang}}, \ and\ \bibinfo {author} {\bibfnamefont {J.~Z.}\
  \bibnamefont {Wu}},\ }\bibfield  {title} {\enquote {\bibinfo {title}
  {Enhancing the capacitive performance of electric double-layer capacitors
  with ionic liquid mixtures},}\ }\href@noop {} {\bibfield  {journal} {\bibinfo
   {journal} {ACS Energy Lett.}\ }\textbf {\bibinfo {volume} {1}},\ \bibinfo
  {pages} {21--26} (\bibinfo {year} {2016})}\BibitemShut {NoStop}%
\bibitem [{\citenamefont {Dudka}\ \emph {et~al.}(2016)\citenamefont {Dudka},
  \citenamefont {Kondrat}, \citenamefont {Kornyshev},\ and\ \citenamefont
  {Oshanin}}]{KoOs16}%
  \BibitemOpen
  \bibfield  {author} {\bibinfo {author} {\bibfnamefont {M.}~\bibnamefont
  {Dudka}}, \bibinfo {author} {\bibfnamefont {S.}~\bibnamefont {Kondrat}},
  \bibinfo {author} {\bibfnamefont {A.}~\bibnamefont {Kornyshev}}, \ and\
  \bibinfo {author} {\bibfnamefont {G.}~\bibnamefont {Oshanin}},\ }\bibfield
  {title} {\enquote {\bibinfo {title} {Phase behavior and structure of a
  superionic liquid in nonpolarized nanoconfinement},}\ }\href@noop {}
  {\bibfield  {journal} {\bibinfo  {journal} {J. Phys.: Condens. Matt.}\
  }\textbf {\bibinfo {volume} {28}},\ \bibinfo {pages} {466007} (\bibinfo
  {year} {2016})}\BibitemShut {NoStop}%
\bibitem [{\citenamefont {Coles}\ \emph {et~al.}(2017)\citenamefont {Coles},
  \citenamefont {Mishin}, \citenamefont {Perkin}, \citenamefont {Fedorov},\
  and\ \citenamefont {Ivanistsev}}]{FeIv17}%
  \BibitemOpen
  \bibfield  {author} {\bibinfo {author} {\bibfnamefont {S.}~\bibnamefont
  {Coles}}, \bibinfo {author} {\bibfnamefont {M.}~\bibnamefont {Mishin}},
  \bibinfo {author} {\bibfnamefont {S.}~\bibnamefont {Perkin}}, \bibinfo
  {author} {\bibfnamefont {M.}~\bibnamefont {Fedorov}}, \ and\ \bibinfo
  {author} {\bibfnamefont {V.}~\bibnamefont {Ivanistsev}},\ }\bibfield  {title}
  {\enquote {\bibinfo {title} {The nanostructure of a lithium glyme solvate
  ionic liquid at electrified interfaces},}\ }\href@noop {} {\bibfield
  {journal} {\bibinfo  {journal} {Phys. Chem. Chem. Phys.}\ }\textbf {\bibinfo
  {volume} {19}},\ \bibinfo {pages} {11004--11010} (\bibinfo {year}
  {2017})}\BibitemShut {NoStop}%
\bibitem [{\citenamefont {Girotto}\ \emph {et~al.}(2017)\citenamefont
  {Girotto}, \citenamefont {Colla}, \citenamefont {{dos Santos}},\ and\
  \citenamefont {Levin}}]{DoLe17}%
  \BibitemOpen
  \bibfield  {author} {\bibinfo {author} {\bibfnamefont {M.}~\bibnamefont
  {Girotto}}, \bibinfo {author} {\bibfnamefont {T.}~\bibnamefont {Colla}},
  \bibinfo {author} {\bibfnamefont {A.}~\bibnamefont {{dos Santos}}}, \ and\
  \bibinfo {author} {\bibfnamefont {Y.}~\bibnamefont {Levin}},\ }\bibfield
  {title} {\enquote {\bibinfo {title} {Lattice model of an ionic liquid at an
  electrified interface},}\ }\href@noop {} {\bibfield  {journal} {\bibinfo
  {journal} {J. Phys. Chem. B}\ }\textbf {\bibinfo {volume} {121}},\ \bibinfo
  {pages} {6408--6415} (\bibinfo {year} {2017})}\BibitemShut {NoStop}%
\bibitem [{\citenamefont {Wong}\ and\ \citenamefont
  {Muthukumar}(2007)}]{WoMu07}%
  \BibitemOpen
  \bibfield  {author} {\bibinfo {author} {\bibfnamefont {C.}~\bibnamefont
  {Wong}}\ and\ \bibinfo {author} {\bibfnamefont {M.}~\bibnamefont
  {Muthukumar}},\ }\bibfield  {title} {\enquote {\bibinfo {title} {Polymer
  capture by electro-osmotic flow of oppositely charged nanopores},}\
  }\href@noop {} {\bibfield  {journal} {\bibinfo  {journal} {J. Chem. Phys.}\
  }\textbf {\bibinfo {volume} {126}},\ \bibinfo {pages} {164903--164905}
  (\bibinfo {year} {2007})}\BibitemShut {NoStop}%
\bibitem [{\citenamefont {Cazade}, \citenamefont {Hartkamp},\ and\
  \citenamefont {Coasne}(2014)}]{HaCo14}%
  \BibitemOpen
  \bibfield  {author} {\bibinfo {author} {\bibfnamefont {P.}~\bibnamefont
  {Cazade}}, \bibinfo {author} {\bibfnamefont {R.}~\bibnamefont {Hartkamp}}, \
  and\ \bibinfo {author} {\bibfnamefont {B.}~\bibnamefont {Coasne}},\
  }\bibfield  {title} {\enquote {\bibinfo {title} {Structure and dynamics of an
  electrolyte confined in charged nanopores},}\ }\href@noop {} {\bibfield
  {journal} {\bibinfo  {journal} {J. Phys. Chem. C}\ }\textbf {\bibinfo
  {volume} {118}},\ \bibinfo {pages} {5061--5072} (\bibinfo {year}
  {2014})}\BibitemShut {NoStop}%
\bibitem [{\citenamefont {Kondrat}\ \emph {et~al.}(2013)\citenamefont
  {Kondrat}, \citenamefont {Georgi}, \citenamefont {Fedorov},\ and\
  \citenamefont {Kornyshev}}]{FeKo11}%
  \BibitemOpen
  \bibfield  {author} {\bibinfo {author} {\bibfnamefont {S.}~\bibnamefont
  {Kondrat}}, \bibinfo {author} {\bibfnamefont {N.}~\bibnamefont {Georgi}},
  \bibinfo {author} {\bibfnamefont {M.}~\bibnamefont {Fedorov}}, \ and\
  \bibinfo {author} {\bibfnamefont {A.}~\bibnamefont {Kornyshev}},\ }\bibfield
  {title} {\enquote {\bibinfo {title} {A superionic state in nano-porous
  double-layer capacitors: Insights from monte carlo simulations},}\
  }\href@noop {} {\bibfield  {journal} {\bibinfo  {journal} {Phys. Chem. Chem.
  Phys.}\ }\textbf {\bibinfo {volume} {13}},\ \bibinfo {pages} {11359--11366}
  (\bibinfo {year} {2013})}\BibitemShut {NoStop}%
\bibitem [{\citenamefont {{dos Santos}}\ and\ \citenamefont
  {Levin}(2015)}]{DoLe15}%
  \BibitemOpen
  \bibfield  {author} {\bibinfo {author} {\bibfnamefont {A.~P.}\ \bibnamefont
  {{dos Santos}}}\ and\ \bibinfo {author} {\bibfnamefont {Y.}~\bibnamefont
  {Levin}},\ }\bibfield  {title} {\enquote {\bibinfo {title} {Electrolytes
  between dielectric charged surfaces: Simulations and theory},}\ }\href@noop
  {} {\bibfield  {journal} {\bibinfo  {journal} {J. Chem. Phys.}\ }\textbf
  {\bibinfo {volume} {142}},\ \bibinfo {pages} {194104} (\bibinfo {year}
  {2015})}\BibitemShut {NoStop}%
\bibitem [{\citenamefont {Colla}\ \emph {et~al.}(2016)\citenamefont {Colla},
  \citenamefont {Girotto}, \citenamefont {{dos Santos}},\ and\ \citenamefont
  {Levin}}]{DoLe16}%
  \BibitemOpen
  \bibfield  {author} {\bibinfo {author} {\bibfnamefont {T.}~\bibnamefont
  {Colla}}, \bibinfo {author} {\bibfnamefont {M.}~\bibnamefont {Girotto}},
  \bibinfo {author} {\bibfnamefont {A.~P.}\ \bibnamefont {{dos Santos}}}, \
  and\ \bibinfo {author} {\bibfnamefont {Y.}~\bibnamefont {Levin}},\ }\bibfield
   {title} {\enquote {\bibinfo {title} {Charge neutrality breakdown in confined
  aqueous electrolytes: Theory and simulation},}\ }\href@noop {} {\bibfield
  {journal} {\bibinfo  {journal} {J. Chem. Phys.}\ }\textbf {\bibinfo {volume}
  {145}},\ \bibinfo {pages} {094704} (\bibinfo {year} {2016})}\BibitemShut
  {NoStop}%
\bibitem [{\citenamefont {Luo}\ \emph {et~al.}(2015)\citenamefont {Luo},
  \citenamefont {Xing}, \citenamefont {Ling}, \citenamefont {Kleinhammes},\
  and\ \citenamefont {Wu}}]{KlWu15}%
  \BibitemOpen
  \bibfield  {author} {\bibinfo {author} {\bibfnamefont {Z.}~\bibnamefont
  {Luo}}, \bibinfo {author} {\bibfnamefont {Y.}~\bibnamefont {Xing}}, \bibinfo
  {author} {\bibfnamefont {Y.}~\bibnamefont {Ling}}, \bibinfo {author}
  {\bibfnamefont {A.}~\bibnamefont {Kleinhammes}}, \ and\ \bibinfo {author}
  {\bibfnamefont {Y.}~\bibnamefont {Wu}},\ }\bibfield  {title} {\enquote
  {\bibinfo {title} {Electroneutrality breakdown and specific ion effects in
  nanoconfined aqueous electrolytes observed by nmr},}\ }\href@noop {}
  {\bibfield  {journal} {\bibinfo  {journal} {Nat. Commun.}\ }\textbf {\bibinfo
  {volume} {6}},\ \bibinfo {pages} {6358--6365} (\bibinfo {year}
  {2015})}\BibitemShut {NoStop}%
\bibitem [{\citenamefont {Linse}\ and\ \citenamefont
  {Lobaskin}(1999)}]{LiLo99}%
  \BibitemOpen
  \bibfield  {author} {\bibinfo {author} {\bibfnamefont {P.}~\bibnamefont
  {Linse}}\ and\ \bibinfo {author} {\bibfnamefont {V.}~\bibnamefont
  {Lobaskin}},\ }\bibfield  {title} {\enquote {\bibinfo {title} {Electrostatic
  attraction and phase separation of like-charged colloidal particles},}\
  }\href@noop {} {\bibfield  {journal} {\bibinfo  {journal} {Phys. Rev. Lett.}\
  }\textbf {\bibinfo {volume} {83}},\ \bibinfo {pages} {4208--4211} (\bibinfo
  {year} {1999})}\BibitemShut {NoStop}%
\bibitem [{\citenamefont {Hatlo}\ and\ \citenamefont {Lue}(2010)}]{HaLu10}%
  \BibitemOpen
  \bibfield  {author} {\bibinfo {author} {\bibfnamefont {M.}~\bibnamefont
  {Hatlo}}\ and\ \bibinfo {author} {\bibfnamefont {L.}~\bibnamefont {Lue}},\
  }\bibfield  {title} {\enquote {\bibinfo {title} {Electrostatic interactions
  of charged bodies from the weak-to the strong-coupling regime},}\ }\href@noop
  {} {\bibfield  {journal} {\bibinfo  {journal} {Euro. Phys. Lett.}\ }\textbf
  {\bibinfo {volume} {89}},\ \bibinfo {pages} {25002} (\bibinfo {year}
  {2010})}\BibitemShut {NoStop}%
\bibitem [{\citenamefont {\v{S}amaj E~Trizac}(2011)}]{SaTr11}%
  \BibitemOpen
  \bibfield  {author} {\bibinfo {author} {\bibfnamefont {L.}~\bibnamefont
  {\v{S}amaj E~Trizac}},\ }\bibfield  {title} {\enquote {\bibinfo {title}
  {Counterions at highly charged interfaces: From one plate to like-charge
  attraction},}\ }\href@noop {} {\bibfield  {journal} {\bibinfo  {journal}
  {Phys. Rev. Lett.}\ }\textbf {\bibinfo {volume} {106}},\ \bibinfo {pages}
  {078301} (\bibinfo {year} {2011})}\BibitemShut {NoStop}%
\bibitem [{\citenamefont {Netz}\ and\ \citenamefont {Orland}(2000)}]{NeOr00}%
  \BibitemOpen
  \bibfield  {author} {\bibinfo {author} {\bibfnamefont {R.}~\bibnamefont
  {Netz}}\ and\ \bibinfo {author} {\bibfnamefont {H.}~\bibnamefont {Orland}},\
  }\bibfield  {title} {\enquote {\bibinfo {title} {Beyond poisson-boltzmann:
  Fluctuation effects and correlation functions},}\ }\href@noop {} {\bibfield
  {journal} {\bibinfo  {journal} {The Euro. Phys. Journ. E}\ }\textbf {\bibinfo
  {volume} {1}},\ \bibinfo {pages} {203--214} (\bibinfo {year}
  {2000})}\BibitemShut {NoStop}%
\bibitem [{\citenamefont {Martin-Molina}\ \emph {et~al.}(2011)\citenamefont
  {Martin-Molina}, \citenamefont {Ibarra-Armenta}, \citenamefont
  {Gonzalez-Tovar}, \citenamefont {Hidalgo-Alvarez},\ and\ \citenamefont
  {Quesada-Perez}}]{AlPe11}%
  \BibitemOpen
  \bibfield  {author} {\bibinfo {author} {\bibfnamefont {A.}~\bibnamefont
  {Martin-Molina}}, \bibinfo {author} {\bibfnamefont {J.~G.}\ \bibnamefont
  {Ibarra-Armenta}}, \bibinfo {author} {\bibfnamefont {E.}~\bibnamefont
  {Gonzalez-Tovar}}, \bibinfo {author} {\bibfnamefont {R.}~\bibnamefont
  {Hidalgo-Alvarez}}, \ and\ \bibinfo {author} {\bibfnamefont {M.}~\bibnamefont
  {Quesada-Perez}},\ }\bibfield  {title} {\enquote {\bibinfo {title} {Monte
  carlo simulations of the electrical double layer forces in the presence of
  divalent electrolyte solutions: Effect of the ion size},}\ }\href@noop {}
  {\bibfield  {journal} {\bibinfo  {journal} {Soft Matter}\ }\textbf {\bibinfo
  {volume} {7}},\ \bibinfo {pages} {1441--1449} (\bibinfo {year}
  {2011})}\BibitemShut {NoStop}%
\bibitem [{\citenamefont {Grosberg}, \citenamefont {Nguyen},\ and\
  \citenamefont {Shklovskii}(2002)}]{NgSh02}%
  \BibitemOpen
  \bibfield  {author} {\bibinfo {author} {\bibfnamefont {A.}~\bibnamefont
  {Grosberg}}, \bibinfo {author} {\bibfnamefont {T.}~\bibnamefont {Nguyen}}, \
  and\ \bibinfo {author} {\bibfnamefont {B.}~\bibnamefont {Shklovskii}},\
  }\bibfield  {title} {\enquote {\bibinfo {title} {Colloquium: The physics of
  charge inversion in chemical and biological systems},}\ }\href@noop {}
  {\bibfield  {journal} {\bibinfo  {journal} {Rev. Mod. Phys.}\ }\textbf
  {\bibinfo {volume} {329}},\ \bibinfo {pages} {329--345} (\bibinfo {year}
  {2002})}\BibitemShut {NoStop}%
\bibitem [{\citenamefont {Wang}\ and\ \citenamefont {Wu}(2017)}]{WaWu17}%
  \BibitemOpen
  \bibfield  {author} {\bibinfo {author} {\bibfnamefont {Z.}~\bibnamefont
  {Wang}}\ and\ \bibinfo {author} {\bibfnamefont {J.}~\bibnamefont {Wu}},\
  }\bibfield  {title} {\enquote {\bibinfo {title} {Ion association at
  discretly-charged dielectric interfaces: Giant charge inversion},}\
  }\href@noop {} {\bibfield  {journal} {\bibinfo  {journal} {J. Chem. Phys.}\
  }\textbf {\bibinfo {volume} {147}},\ \bibinfo {pages} {024703} (\bibinfo
  {year} {2017})}\BibitemShut {NoStop}%
\bibitem [{\citenamefont {Levin}(2002)}]{Lev02}%
  \BibitemOpen
  \bibfield  {author} {\bibinfo {author} {\bibfnamefont {Y.}~\bibnamefont
  {Levin}},\ }\bibfield  {title} {\enquote {\bibinfo {title} {Electrostatic
  correlations: from plasma to biology},}\ }\href@noop {} {\bibfield  {journal}
  {\bibinfo  {journal} {Rep. Prog. Phys.}\ }\textbf {\bibinfo {volume} {65}},\
  \bibinfo {pages} {1577} (\bibinfo {year} {2002})}\BibitemShut {NoStop}%
\bibitem [{\citenamefont {Yeh}\ and\ \citenamefont {Berkowitz}(1999)}]{YeBe99}%
  \BibitemOpen
  \bibfield  {author} {\bibinfo {author} {\bibfnamefont {I.~C.}\ \bibnamefont
  {Yeh}}\ and\ \bibinfo {author} {\bibfnamefont {M.~L.}\ \bibnamefont
  {Berkowitz}},\ }\bibfield  {title} {\enquote {\bibinfo {title} {Ewald
  summation for systems with slab geometry},}\ }\href@noop {} {\bibfield
  {journal} {\bibinfo  {journal} {J. Chem. Phys.}\ }\textbf {\bibinfo {volume}
  {111}},\ \bibinfo {pages} {3155--3162} (\bibinfo {year} {1999})}\BibitemShut
  {NoStop}%
\bibitem [{\citenamefont {Kawata}\ and\ \citenamefont {Mikami}(2001)}]{KaMi01}%
  \BibitemOpen
  \bibfield  {author} {\bibinfo {author} {\bibfnamefont {M.}~\bibnamefont
  {Kawata}}\ and\ \bibinfo {author} {\bibfnamefont {M.}~\bibnamefont
  {Mikami}},\ }\bibfield  {title} {\enquote {\bibinfo {title} {Rapid
  calculation of two-dimensional ewald summation},}\ }\href@noop {} {\bibfield
  {journal} {\bibinfo  {journal} {Chem. Phys. Lett.}\ }\textbf {\bibinfo
  {volume} {340}},\ \bibinfo {pages} {157--164} (\bibinfo {year}
  {2001})}\BibitemShut {NoStop}%
\bibitem [{\citenamefont {Arnold}, \citenamefont {{de Joannis}},\ and\
  \citenamefont {Holm}(2002)}]{ArDe02}%
  \BibitemOpen
  \bibfield  {author} {\bibinfo {author} {\bibfnamefont {A.}~\bibnamefont
  {Arnold}}, \bibinfo {author} {\bibfnamefont {J.}~\bibnamefont {{de
  Joannis}}}, \ and\ \bibinfo {author} {\bibfnamefont {C.}~\bibnamefont
  {Holm}},\ }\bibfield  {title} {\enquote {\bibinfo {title} {Electrostatics in
  periodic slab geometries. i},}\ }\href@noop {} {\bibfield  {journal}
  {\bibinfo  {journal} {J. Chem. Phys.}\ }\textbf {\bibinfo {volume} {117}},\
  \bibinfo {pages} {2496--2502} (\bibinfo {year} {2002})}\BibitemShut {NoStop}%
\bibitem [{\citenamefont {Nagy}, \citenamefont {Henderson},\ and\ \citenamefont
  {Boda}(2011)}]{NaHe11}%
  \BibitemOpen
  \bibfield  {author} {\bibinfo {author} {\bibfnamefont {T.}~\bibnamefont
  {Nagy}}, \bibinfo {author} {\bibfnamefont {D.}~\bibnamefont {Henderson}}, \
  and\ \bibinfo {author} {\bibfnamefont {D.}~\bibnamefont {Boda}},\ }\bibfield
  {title} {\enquote {\bibinfo {title} {Simulation of an electrical double layer
  model with a low dielectric layer between the electrode and the
  electrolyte},}\ }\href@noop {} {\bibfield  {journal} {\bibinfo  {journal} {J.
  Phys. Chem. B}\ }\textbf {\bibinfo {volume} {115}},\ \bibinfo {pages}
  {11409--11419} (\bibinfo {year} {2011})}\BibitemShut {NoStop}%
\bibitem [{\citenamefont {Wang}\ and\ \citenamefont {Ma}(2016)}]{WaMa16}%
  \BibitemOpen
  \bibfield  {author} {\bibinfo {author} {\bibfnamefont {Z.-Y.}\ \bibnamefont
  {Wang}}\ and\ \bibinfo {author} {\bibfnamefont {Z.}~\bibnamefont {Ma}},\
  }\bibfield  {title} {\enquote {\bibinfo {title} {Examining the contributions
  of image-charge forces to charge reversal: Discrete versus continuum modeling
  of surface charges},}\ }\href@noop {} {\bibfield  {journal} {\bibinfo
  {journal} {J. Chem. Theory Comput.}\ }\textbf {\bibinfo {volume} {12}},\
  \bibinfo {pages} {2880--2888} (\bibinfo {year} {2016})}\BibitemShut {NoStop}%
\bibitem [{\citenamefont {Girotto}, \citenamefont {{dos Santos}},\ and\
  \citenamefont {Levin}(2016)}]{GiDo16}%
  \BibitemOpen
  \bibfield  {author} {\bibinfo {author} {\bibfnamefont {M.}~\bibnamefont
  {Girotto}}, \bibinfo {author} {\bibfnamefont {A.~P.}\ \bibnamefont {{dos
  Santos}}}, \ and\ \bibinfo {author} {\bibfnamefont {Y.}~\bibnamefont
  {Levin}},\ }\bibfield  {title} {\enquote {\bibinfo {title} {Interaction of
  charged colloidal particles at the water-air interface},}\ }\href@noop {}
  {\bibfield  {journal} {\bibinfo  {journal} {J. Phys. Chem. B}\ }\textbf
  {\bibinfo {volume} {120}},\ \bibinfo {pages} {5817--5822} (\bibinfo {year}
  {2016})}\BibitemShut {NoStop}%
\bibitem [{\citenamefont {{Guerrero-García}}, \citenamefont {Jing},\ and\
  \citenamefont {{de la Cruz}}(2013)}]{JiCr13}%
  \BibitemOpen
  \bibfield  {author} {\bibinfo {author} {\bibfnamefont {G.}~\bibnamefont
  {{Guerrero-García}}}, \bibinfo {author} {\bibfnamefont {Y.}~\bibnamefont
  {Jing}}, \ and\ \bibinfo {author} {\bibfnamefont {M.}~\bibnamefont {{de la
  Cruz}}},\ }\bibfield  {title} {\enquote {\bibinfo {title} {Enhancing and
  reversing the electric field at the oil–water interface with
  size-asymmetric monovalent ions},}\ }\href@noop {} {\bibfield  {journal}
  {\bibinfo  {journal} {Soft Matter}\ }\textbf {\bibinfo {volume} {9}},\
  \bibinfo {pages} {6046--6052} (\bibinfo {year} {2013})}\BibitemShut {NoStop}%
\bibitem [{\citenamefont {Bakhshandeh}, \citenamefont {{dos Santos}},\ and\
  \citenamefont {Levin}(2011)}]{BaDo11}%
  \BibitemOpen
  \bibfield  {author} {\bibinfo {author} {\bibfnamefont {A.}~\bibnamefont
  {Bakhshandeh}}, \bibinfo {author} {\bibfnamefont {A.~P.}\ \bibnamefont {{dos
  Santos}}}, \ and\ \bibinfo {author} {\bibfnamefont {Y.}~\bibnamefont
  {Levin}},\ }\bibfield  {title} {\enquote {\bibinfo {title} {Weak and strong
  coupling theories for polarizable colloids and nanoparticles},}\ }\href@noop
  {} {\bibfield  {journal} {\bibinfo  {journal} {Phys. Rev. Lett.}\ }\textbf
  {\bibinfo {volume} {107}},\ \bibinfo {pages} {107801} (\bibinfo {year}
  {2011})}\BibitemShut {NoStop}%
\bibitem [{\citenamefont {{dos Santos}}, \citenamefont {Bakhshandeh},\ and\
  \citenamefont {Levin}(2011)}]{DoBa11}%
  \BibitemOpen
  \bibfield  {author} {\bibinfo {author} {\bibfnamefont {A.~P.}\ \bibnamefont
  {{dos Santos}}}, \bibinfo {author} {\bibfnamefont {A.}~\bibnamefont
  {Bakhshandeh}}, \ and\ \bibinfo {author} {\bibfnamefont {Y.}~\bibnamefont
  {Levin}},\ }\bibfield  {title} {\enquote {\bibinfo {title} {Effects of the
  dielectric discontinuity on the counterion distribution in a colloidal
  suspension},}\ }\href@noop {} {\bibfield  {journal} {\bibinfo  {journal} {J.
  Chem. Phys.}\ }\textbf {\bibinfo {volume} {135}},\ \bibinfo {pages} {044124}
  (\bibinfo {year} {2011})}\BibitemShut {NoStop}%
\bibitem [{\citenamefont {Diehl}, \citenamefont {{dos Santos}},\ and\
  \citenamefont {Levin}(2012)}]{DiDo12}%
  \BibitemOpen
  \bibfield  {author} {\bibinfo {author} {\bibfnamefont {A.}~\bibnamefont
  {Diehl}}, \bibinfo {author} {\bibfnamefont {A.~P.}\ \bibnamefont {{dos
  Santos}}}, \ and\ \bibinfo {author} {\bibfnamefont {Y.}~\bibnamefont
  {Levin}},\ }\bibfield  {title} {\enquote {\bibinfo {title} {Surface tension
  of electrolyte-air interface: A monte carlo study},}\ }\href@noop {}
  {\bibfield  {journal} {\bibinfo  {journal} {J. Phys.: Condens. Matt.}\
  }\textbf {\bibinfo {volume} {24}},\ \bibinfo {pages} {284115} (\bibinfo
  {year} {2012})}\BibitemShut {NoStop}%
\bibitem [{\citenamefont {Siepmann}\ and\ \citenamefont
  {Sprik}(1995)}]{SiSp95}%
  \BibitemOpen
  \bibfield  {author} {\bibinfo {author} {\bibfnamefont {J.~I.}\ \bibnamefont
  {Siepmann}}\ and\ \bibinfo {author} {\bibfnamefont {M.}~\bibnamefont
  {Sprik}},\ }\bibfield  {title} {\enquote {\bibinfo {title} {Influence of
  surface topology and electrostatic potential on water/electrode systems},}\
  }\href@noop {} {\bibfield  {journal} {\bibinfo  {journal} {J. Chem. Phys.}\
  }\textbf {\bibinfo {volume} {102}},\ \bibinfo {pages} {511--524} (\bibinfo
  {year} {1995})}\BibitemShut {NoStop}%
\bibitem [{\citenamefont {{dos Santos}}, \citenamefont {Girotto},\ and\
  \citenamefont {Levin}(2016{\natexlab{a}})}]{GiLe16}%
  \BibitemOpen
  \bibfield  {author} {\bibinfo {author} {\bibfnamefont {A.~P.}\ \bibnamefont
  {{dos Santos}}}, \bibinfo {author} {\bibfnamefont {M.}~\bibnamefont
  {Girotto}}, \ and\ \bibinfo {author} {\bibfnamefont {Y.}~\bibnamefont
  {Levin}},\ }\bibfield  {title} {\enquote {\bibinfo {title} {Simulations of
  polyelectrolyte adsorption to a dielectric like-charged surface},}\
  }\href@noop {} {\bibfield  {journal} {\bibinfo  {journal} {J. Phys. Chem. B}\
  }\textbf {\bibinfo {volume} {120}},\ \bibinfo {pages} {10387--10393}
  (\bibinfo {year} {2016}{\natexlab{a}})}\BibitemShut {NoStop}%
\bibitem [{\citenamefont {Zwanikken}\ and\ \citenamefont {{de la
  Cruz}}(2013)}]{ZwCr13}%
  \BibitemOpen
  \bibfield  {author} {\bibinfo {author} {\bibfnamefont {J.}~\bibnamefont
  {Zwanikken}}\ and\ \bibinfo {author} {\bibfnamefont {M.}~\bibnamefont {{de la
  Cruz}}},\ }\bibfield  {title} {\enquote {\bibinfo {title} {Tunable soft
  structure in charged fluids confined by dielectric interfaces},}\ }\href@noop
  {} {\bibfield  {journal} {\bibinfo  {journal} {Proc. Natl. Acad. Sci.
  U.S.A.}\ }\textbf {\bibinfo {volume} {110}},\ \bibinfo {pages} {5301--5308}
  (\bibinfo {year} {2013})}\BibitemShut {NoStop}%
\bibitem [{\citenamefont {Jadhao}, \citenamefont {Solis},\ and\ \citenamefont
  {{de la Cruz}}(2012)}]{SoCr12}%
  \BibitemOpen
  \bibfield  {author} {\bibinfo {author} {\bibfnamefont {V.}~\bibnamefont
  {Jadhao}}, \bibinfo {author} {\bibfnamefont {F.}~\bibnamefont {Solis}}, \
  and\ \bibinfo {author} {\bibfnamefont {M.}~\bibnamefont {{de la Cruz}}},\
  }\bibfield  {title} {\enquote {\bibinfo {title} {Simulation of charged
  systems in heterogeneous dielectric media via a true energy functional},}\
  }\href@noop {} {\bibfield  {journal} {\bibinfo  {journal} {Phys. Rev. Lett.}\
  }\textbf {\bibinfo {volume} {109}},\ \bibinfo {pages} {223905} (\bibinfo
  {year} {2012})}\BibitemShut {NoStop}%
\bibitem [{\citenamefont {Limmer}\ \emph {et~al.}(2013)\citenamefont {Limmer},
  \citenamefont {Merlet}, \citenamefont {Sallane}, \citenamefont {Chandler},
  \citenamefont {Madden}, \citenamefont {{van Roij}},\ and\ \citenamefont
  {Rotenberg}}]{RoRo13}%
  \BibitemOpen
  \bibfield  {author} {\bibinfo {author} {\bibfnamefont {D.~T.}\ \bibnamefont
  {Limmer}}, \bibinfo {author} {\bibfnamefont {C.}~\bibnamefont {Merlet}},
  \bibinfo {author} {\bibfnamefont {M.}~\bibnamefont {Sallane}}, \bibinfo
  {author} {\bibfnamefont {D.}~\bibnamefont {Chandler}}, \bibinfo {author}
  {\bibfnamefont {P.~A.}\ \bibnamefont {Madden}}, \bibinfo {author}
  {\bibfnamefont {R.}~\bibnamefont {{van Roij}}}, \ and\ \bibinfo {author}
  {\bibfnamefont {B.}~\bibnamefont {Rotenberg}},\ }\bibfield  {title} {\enquote
  {\bibinfo {title} {Charge fluctuations in nanoscale capacitors},}\
  }\href@noop {} {\bibfield  {journal} {\bibinfo  {journal} {Phys. Rev. Lett.}\
  }\textbf {\bibinfo {volume} {111}},\ \bibinfo {pages} {106102--106106}
  (\bibinfo {year} {2013})}\BibitemShut {NoStop}%
\bibitem [{\citenamefont {Merlet}\ \emph {et~al.}(2014)\citenamefont {Merlet},
  \citenamefont {Limmer}, \citenamefont {Salanne}, \citenamefont {{van Roij}},
  \citenamefont {Madden}, \citenamefont {Chandler},\ and\ \citenamefont
  {Rotenberg}}]{ChRo14}%
  \BibitemOpen
  \bibfield  {author} {\bibinfo {author} {\bibfnamefont {C.}~\bibnamefont
  {Merlet}}, \bibinfo {author} {\bibfnamefont {D.~T.}\ \bibnamefont {Limmer}},
  \bibinfo {author} {\bibfnamefont {M.}~\bibnamefont {Salanne}}, \bibinfo
  {author} {\bibfnamefont {R.}~\bibnamefont {{van Roij}}}, \bibinfo {author}
  {\bibfnamefont {P.~A.}\ \bibnamefont {Madden}}, \bibinfo {author}
  {\bibfnamefont {D.}~\bibnamefont {Chandler}}, \ and\ \bibinfo {author}
  {\bibfnamefont {B.}~\bibnamefont {Rotenberg}},\ }\bibfield  {title} {\enquote
  {\bibinfo {title} {The electric double layer has a life of its own},}\
  }\href@noop {} {\bibfield  {journal} {\bibinfo  {journal} {J. Phys. Chem. C}\
  }\textbf {\bibinfo {volume} {118}},\ \bibinfo {pages} {18291--18298}
  (\bibinfo {year} {2014})}\BibitemShut {NoStop}%
\bibitem [{\citenamefont {Jing}\ \emph {et~al.}(2015)\citenamefont {Jing},
  \citenamefont {Jadhao}, \citenamefont {Zwanikken},\ and\ \citenamefont {{de
  La Cruz}}}]{ZwLa15}%
  \BibitemOpen
  \bibfield  {author} {\bibinfo {author} {\bibfnamefont {Y.}~\bibnamefont
  {Jing}}, \bibinfo {author} {\bibfnamefont {V.}~\bibnamefont {Jadhao}},
  \bibinfo {author} {\bibfnamefont {J.}~\bibnamefont {Zwanikken}}, \ and\
  \bibinfo {author} {\bibfnamefont {M.}~\bibnamefont {{de La Cruz}}},\
  }\bibfield  {title} {\enquote {\bibinfo {title} {Ionic structure in liquids
  confined by dielectric surfaces},}\ }\href@noop {} {\bibfield  {journal}
  {\bibinfo  {journal} {J. Chem. Phys.}\ }\textbf {\bibinfo {volume} {143}},\
  \bibinfo {pages} {194508} (\bibinfo {year} {2015})}\BibitemShut {NoStop}%
\bibitem [{\citenamefont {Boda}\ \emph {et~al.}(2004)\citenamefont {Boda},
  \citenamefont {Gillespie}, \citenamefont {Nonner}, \citenamefont
  {Henderson},\ and\ \citenamefont {Eisenberg}}]{HeEi04}%
  \BibitemOpen
  \bibfield  {author} {\bibinfo {author} {\bibfnamefont {D.}~\bibnamefont
  {Boda}}, \bibinfo {author} {\bibfnamefont {D.}~\bibnamefont {Gillespie}},
  \bibinfo {author} {\bibfnamefont {W.}~\bibnamefont {Nonner}}, \bibinfo
  {author} {\bibfnamefont {D.}~\bibnamefont {Henderson}}, \ and\ \bibinfo
  {author} {\bibfnamefont {B.}~\bibnamefont {Eisenberg}},\ }\bibfield  {title}
  {\enquote {\bibinfo {title} {Computing induced charges in inhomogeneous
  dielectric media: Application in a monte carlo simulation of complex ionic
  systems},}\ }\href@noop {} {\bibfield  {journal} {\bibinfo  {journal} {Phys.
  Rev. E}\ }\textbf {\bibinfo {volume} {69}},\ \bibinfo {pages} {046702}
  (\bibinfo {year} {2004})}\BibitemShut {NoStop}%
\bibitem [{\citenamefont {Gan}\ \emph {et~al.}(2015)\citenamefont {Gan},
  \citenamefont {Wu}, \citenamefont {Barros}, \citenamefont {Xu},\ and\
  \citenamefont {Luijten}}]{XuLu15}%
  \BibitemOpen
  \bibfield  {author} {\bibinfo {author} {\bibfnamefont {Z.}~\bibnamefont
  {Gan}}, \bibinfo {author} {\bibfnamefont {H.}~\bibnamefont {Wu}}, \bibinfo
  {author} {\bibfnamefont {K.}~\bibnamefont {Barros}}, \bibinfo {author}
  {\bibfnamefont {Z.}~\bibnamefont {Xu}}, \ and\ \bibinfo {author}
  {\bibfnamefont {E.}~\bibnamefont {Luijten}},\ }\bibfield  {title} {\enquote
  {\bibinfo {title} {Comparison of efficient techniques for the simulation of
  dielectric objects in electrolytes},}\ }\href@noop {} {\bibfield  {journal}
  {\bibinfo  {journal} {J. Comp. Phys.}\ }\textbf {\bibinfo {volume} {291}},\
  \bibinfo {pages} {317--333} (\bibinfo {year} {2015})}\BibitemShut {NoStop}%
\bibitem [{\citenamefont {Tyagi}\ \emph {et~al.}(2010)\citenamefont {Tyagi},
  \citenamefont {Süzen}, \citenamefont {Sega}, \citenamefont {Barbosa},
  \citenamefont {Kantarovitch},\ and\ \citenamefont {Holm}}]{KaHo10}%
  \BibitemOpen
  \bibfield  {author} {\bibinfo {author} {\bibfnamefont {S.}~\bibnamefont
  {Tyagi}}, \bibinfo {author} {\bibfnamefont {M.}~\bibnamefont {Süzen}},
  \bibinfo {author} {\bibfnamefont {M.}~\bibnamefont {Sega}}, \bibinfo {author}
  {\bibfnamefont {M.}~\bibnamefont {Barbosa}}, \bibinfo {author} {\bibfnamefont
  {S.}~\bibnamefont {Kantarovitch}}, \ and\ \bibinfo {author} {\bibfnamefont
  {C.}~\bibnamefont {Holm}},\ }\bibfield  {title} {\enquote {\bibinfo {title}
  {An iterative, fast, linear-scaling method for computing induced charges on
  arbitrary dielectric boundaries},}\ }\href@noop {} {\bibfield  {journal}
  {\bibinfo  {journal} {J. Chem. Phys.}\ }\textbf {\bibinfo {volume} {132}},\
  \bibinfo {pages} {154112} (\bibinfo {year} {2010})}\BibitemShut {NoStop}%
\bibitem [{\citenamefont {Girotto}, \citenamefont {{dos Santos}},\ and\
  \citenamefont {Levin}(2017)}]{DoLe172}%
  \BibitemOpen
  \bibfield  {author} {\bibinfo {author} {\bibfnamefont {M.}~\bibnamefont
  {Girotto}}, \bibinfo {author} {\bibfnamefont {A.~P.}\ \bibnamefont {{dos
  Santos}}}, \ and\ \bibinfo {author} {\bibfnamefont {Y.}~\bibnamefont
  {Levin}},\ }\bibfield  {title} {\enquote {\bibinfo {title} {Simulations of
  ionic liquids confined by metal electrodes using periodic green functions},}\
  }\href@noop {} {\bibfield  {journal} {\bibinfo  {journal} {J. Chem. Phys.}\
  }\textbf {\bibinfo {volume} {147}},\ \bibinfo {pages} {074109} (\bibinfo
  {year} {2017})}\BibitemShut {NoStop}%
\bibitem [{\citenamefont {Arnold}\ \emph {et~al.}(2013)\citenamefont {Arnold},
  \citenamefont {Breitsprecher}, \citenamefont {Fahrenberger}, \citenamefont
  {Kesselheim}, \citenamefont {Lenz},\ and\ \citenamefont {Holm}}]{LeHo13}%
  \BibitemOpen
  \bibfield  {author} {\bibinfo {author} {\bibfnamefont {A.}~\bibnamefont
  {Arnold}}, \bibinfo {author} {\bibfnamefont {K.}~\bibnamefont
  {Breitsprecher}}, \bibinfo {author} {\bibfnamefont {F.}~\bibnamefont
  {Fahrenberger}}, \bibinfo {author} {\bibfnamefont {S.}~\bibnamefont
  {Kesselheim}}, \bibinfo {author} {\bibfnamefont {O.}~\bibnamefont {Lenz}}, \
  and\ \bibinfo {author} {\bibfnamefont {C.}~\bibnamefont {Holm}},\ }\bibfield
  {title} {\enquote {\bibinfo {title} {Efficient algorithms for electrostatic
  interactions including dielectric contrasts},}\ }\href@noop {} {\bibfield
  {journal} {\bibinfo  {journal} {Entropy}\ }\textbf {\bibinfo {volume} {15}},\
  \bibinfo {pages} {4569--4588} (\bibinfo {year} {2013})}\BibitemShut {NoStop}%
\bibitem [{\citenamefont {Tyagi}, \citenamefont {Arnold},\ and\ \citenamefont
  {Holm}(2008)}]{ArHo08}%
  \BibitemOpen
  \bibfield  {author} {\bibinfo {author} {\bibfnamefont {S.}~\bibnamefont
  {Tyagi}}, \bibinfo {author} {\bibfnamefont {A.}~\bibnamefont {Arnold}}, \
  and\ \bibinfo {author} {\bibfnamefont {C.}~\bibnamefont {Holm}},\ }\bibfield
  {title} {\enquote {\bibinfo {title} {Electrostatic layer correction with
  image charges: A linear scaling method to treat slab 2d+h systems with
  dielectric interfaces},}\ }\href@noop {} {\bibfield  {journal} {\bibinfo
  {journal} {J. Chem. Phys.}\ }\textbf {\bibinfo {volume} {129}},\ \bibinfo
  {pages} {204102} (\bibinfo {year} {2008})}\BibitemShut {NoStop}%
\bibitem [{\citenamefont {Tyagi}, \citenamefont {Arnold},\ and\ \citenamefont
  {Holm}(2007)}]{ArHo07}%
  \BibitemOpen
  \bibfield  {author} {\bibinfo {author} {\bibfnamefont {S.}~\bibnamefont
  {Tyagi}}, \bibinfo {author} {\bibfnamefont {A.}~\bibnamefont {Arnold}}, \
  and\ \bibinfo {author} {\bibfnamefont {C.}~\bibnamefont {Holm}},\ }\bibfield
  {title} {\enquote {\bibinfo {title} {Icmmm2d: An accurate method to include
  planar dielectric interfaces via image charge summation},}\ }\href@noop {}
  {\bibfield  {journal} {\bibinfo  {journal} {J. Chem. Phys.}\ }\textbf
  {\bibinfo {volume} {127}},\ \bibinfo {pages} {154723} (\bibinfo {year}
  {2007})}\BibitemShut {NoStop}%
\bibitem [{\citenamefont {{J. D. Jackson}}(1999)}]{Jac99}%
  \BibitemOpen
  \bibfield  {author} {\bibinfo {author} {\bibnamefont {{J. D. Jackson}}},\
  }\href@noop {} {\emph {\bibinfo {title} {Classical Electrodynamics}}}\
  (\bibinfo  {publisher} {Wyley},\ \bibinfo {year} {1999})\ pp.\ \bibinfo
  {pages} {140--141}\BibitemShut {NoStop}%
\bibitem [{\citenamefont {{dos Santos}}, \citenamefont {Girotto},\ and\
  \citenamefont {Levin}(2016{\natexlab{b}})}]{DoGi16}%
  \BibitemOpen
  \bibfield  {author} {\bibinfo {author} {\bibfnamefont {A.~P.}\ \bibnamefont
  {{dos Santos}}}, \bibinfo {author} {\bibfnamefont {M.}~\bibnamefont
  {Girotto}}, \ and\ \bibinfo {author} {\bibfnamefont {Y.}~\bibnamefont
  {Levin}},\ }\bibfield  {title} {\enquote {\bibinfo {title} {Simulations of
  coulomb systems with slab geometry using an efficient 3d ewald summation
  method},}\ }\href@noop {} {\bibfield  {journal} {\bibinfo  {journal} {J.
  Chem. Phys.}\ }\textbf {\bibinfo {volume} {144}},\ \bibinfo {pages} {144103}
  (\bibinfo {year} {2016}{\natexlab{b}})}\BibitemShut {NoStop}%
\bibitem [{\citenamefont {Metropolis}\ \emph {et~al.}(1953)\citenamefont
  {Metropolis}, \citenamefont {Rosenbluth}, \citenamefont {Rosenbluth},
  \citenamefont {Teller},\ and\ \citenamefont {Teller}}]{NiTe53}%
  \BibitemOpen
  \bibfield  {author} {\bibinfo {author} {\bibfnamefont {N.}~\bibnamefont
  {Metropolis}}, \bibinfo {author} {\bibfnamefont {A.~W.}\ \bibnamefont
  {Rosenbluth}}, \bibinfo {author} {\bibfnamefont {M.~N.}\ \bibnamefont
  {Rosenbluth}}, \bibinfo {author} {\bibfnamefont {A.~H.}\ \bibnamefont
  {Teller}}, \ and\ \bibinfo {author} {\bibfnamefont {E.}~\bibnamefont
  {Teller}},\ }\bibfield  {title} {\enquote {\bibinfo {title} {Equation of
  state calculations by fast computing machines},}\ }\href@noop {} {\bibfield
  {journal} {\bibinfo  {journal} {J. Chem. Phys.}\ }\textbf {\bibinfo {volume}
  {21}},\ \bibinfo {pages} {1087--1092} (\bibinfo {year} {1953})}\BibitemShut
  {NoStop}%
\bibitem [{\citenamefont {Hautman}\ and\ \citenamefont {Klein}(1992)}]{Kl92}%
  \BibitemOpen
  \bibfield  {author} {\bibinfo {author} {\bibfnamefont {J.}~\bibnamefont
  {Hautman}}\ and\ \bibinfo {author} {\bibfnamefont {M.~L.}\ \bibnamefont
  {Klein}},\ }\bibfield  {title} {\enquote {\bibinfo {title} {An ewald
  summation method for planar surfaces and interfaces},}\ }\href@noop {}
  {\bibfield  {journal} {\bibinfo  {journal} {Mol. Phys.}\ }\textbf {\bibinfo
  {volume} {75}},\ \bibinfo {pages} {379--395} (\bibinfo {year}
  {1992})}\BibitemShut {NoStop}%
\bibitem [{\citenamefont {Spohr}(1997)}]{Sp97}%
  \BibitemOpen
  \bibfield  {author} {\bibinfo {author} {\bibfnamefont {E.}~\bibnamefont
  {Spohr}},\ }\bibfield  {title} {\enquote {\bibinfo {title} {Effect of
  electrostatic boundary conditions and system size on the interfacial
  properties of water and aqueous solutions},}\ }\href@noop {} {\bibfield
  {journal} {\bibinfo  {journal} {J. Chem. Phys.}\ }\textbf {\bibinfo {volume}
  {107}},\ \bibinfo {pages} {6342--6348} (\bibinfo {year} {1997})}\BibitemShut
  {NoStop}%
\bibitem [{\citenamefont {Lekner}(1989)}]{Lek89}%
  \BibitemOpen
  \bibfield  {author} {\bibinfo {author} {\bibfnamefont {J.}~\bibnamefont
  {Lekner}},\ }\bibfield  {title} {\enquote {\bibinfo {title} {Summation of
  dipolar fields in simulated liquid-vapour interfaces},}\ }\href@noop {}
  {\bibfield  {journal} {\bibinfo  {journal} {Physica A}\ }\textbf {\bibinfo
  {volume} {157}},\ \bibinfo {pages} {826--838} (\bibinfo {year}
  {1989})}\BibitemShut {NoStop}%
\bibitem [{\citenamefont {Sperb}(1994)}]{Spe94}%
  \BibitemOpen
  \bibfield  {author} {\bibinfo {author} {\bibfnamefont {R.}~\bibnamefont
  {Sperb}},\ }\bibfield  {title} {\enquote {\bibinfo {title} {Extension and
  simple proof of lekner's summation formula for coulomb forces},}\ }\href@noop
  {} {\bibfield  {journal} {\bibinfo  {journal} {Mol. Phys.}\ }\textbf
  {\bibinfo {volume} {13}},\ \bibinfo {pages} {189--193} (\bibinfo {year}
  {1994})}\BibitemShut {NoStop}%
\bibitem [{\citenamefont {Arnold}\ and\ \citenamefont {Holm}(2002)}]{ArHo02}%
  \BibitemOpen
  \bibfield  {author} {\bibinfo {author} {\bibfnamefont {A.}~\bibnamefont
  {Arnold}}\ and\ \bibinfo {author} {\bibfnamefont {C.}~\bibnamefont {Holm}},\
  }\bibfield  {title} {\enquote {\bibinfo {title} {A novel method for
  calculating electrostatic interactions in 2d periodic slab geometries},}\
  }\href@noop {} {\bibfield  {journal} {\bibinfo  {journal} {Chem. Phys.
  Lett.}\ }\textbf {\bibinfo {volume} {354}},\ \bibinfo {pages} {324--330}
  (\bibinfo {year} {2002})}\BibitemShut {NoStop}%
\bibitem [{\citenamefont {Mináry}\ \emph {et~al.}(2002)\citenamefont
  {Mináry}, \citenamefont {Tuckerman}, \citenamefont {Pihakari},\ and\
  \citenamefont {Martyna}}]{PiMa02}%
  \BibitemOpen
  \bibfield  {author} {\bibinfo {author} {\bibfnamefont {P.}~\bibnamefont
  {Mináry}}, \bibinfo {author} {\bibfnamefont {M.}~\bibnamefont {Tuckerman}},
  \bibinfo {author} {\bibfnamefont {K.}~\bibnamefont {Pihakari}}, \ and\
  \bibinfo {author} {\bibfnamefont {G.}~\bibnamefont {Martyna}},\ }\bibfield
  {title} {\enquote {\bibinfo {title} {A new reciprocal space based treatment
  of long range interactions on surfaces},}\ }\href@noop {} {\bibfield
  {journal} {\bibinfo  {journal} {J. Chem. Phys.}\ }\textbf {\bibinfo {volume}
  {116}},\ \bibinfo {pages} {5351--5362} (\bibinfo {year} {2002})}\BibitemShut
  {NoStop}%
\bibitem [{\citenamefont {Smith}(1981)}]{Sm81}%
  \BibitemOpen
  \bibfield  {author} {\bibinfo {author} {\bibfnamefont {E.~R.}\ \bibnamefont
  {Smith}},\ }\bibfield  {title} {\enquote {\bibinfo {title} {Electrostatic
  energy in ionic crystals},}\ }\href@noop {} {\bibfield  {journal} {\bibinfo
  {journal} {Proc. R. Soc. Lond. A}\ }\textbf {\bibinfo {volume} {375}},\
  \bibinfo {pages} {475--505} (\bibinfo {year} {1981})}\BibitemShut {NoStop}%
\end{thebibliography}%

\end{document}